\documentclass[10pt,final,journal,letterpaper,oneside,twocolumn]{IEEEtran}
\pdfoutput=1
\usepackage{amsmath,amssymb}
\usepackage[abs]{overpic}
\usepackage[pdftex]{pict2e}
\usepackage[usenames,dvipsnames]{color}

\newcommand{\cM}{\mathcal{M}}
\newcommand{\haihun}{\mathchar`-}
\newcommand{\vep}{\varepsilon}
\newcommand{\phiBr}{\phi(\rho|W_B,q,r)}
\newcommand{\phiEr}{\phi(-\rho|W_E,q,r)}
\newcommand{\Fc}{F_{c}(q, R_B, R_E)}
\newcommand{\Hc}{H_{c}(q, R_E)}
\newcommand{\non}{\nonumber}

\newcounter{MYtempeqncnt}
\title{Numerical Study on Secrecy Capacity and Code Length Dependence of the Performances in Optical Wiretap Channels}

\author{H.~Endo, T.~S.~Han,~\IEEEmembership{Life~Fellow,~IEEE}, T.~Aoki, M.~Sasaki%
\thanks{This work was supported by the Council for Science, Technology, and Innovation (Cabinet Office, Government of Japan) through the ImPACT Program.}%
\thanks{H.~Endo, T.~S.~Han, and M.~Sasaki are with the Quantum ICT Laboratory, National Institute of Information and Communications Technology, Koganei, 184-8795, Japan
(e-mail: h-endo@nict.go.jp; tshan@nict.go.jp; psasaki@nict.go.jp).}%
\thanks{H.~Endo and T.~Aoki are with the Department of Applied Physics, Waseda University, Shinjuku, 169-8050, Japan
(e-mail: h-endo-1212@ruri.waseda.jp; takao@waseda.jp).}}

\begin{document}
\bibliographystyle{IEEEtran}
\maketitle
\begin{abstract}
  Secrecy issues of free-space optical links realizing information theoretically secure communications as well as high transmission rates are discussed.  
  We numerically study secrecy communication rates of optical wiretap channel based on on-off keying modulation under typical conditions met in satellite-ground links.  
  It is shown that under reasonable degraded conditions on a wiretapper, 
  information theoretically secure communications should be possible in a much wider distance range than a range limit of quantum key distribution, 
  enabling secure optical links between geostationary earth orbit satellites and ground stations with currently available technologies.  
  We also provide the upper bounds on the decoding error probability and the leaked information to estimate a necessary code length for given required levels of performances.  
  This result ensures that a reasonable length wiretap channel code for our proposed scheme must exist.
\end{abstract}
\begin{IEEEkeywords}
Physical layer security, free space optical communication, secrecy capacity, finite-length analysis.
\end{IEEEkeywords}

\section{Introduction}
  Free-space optical (FSO) communication is a promising technology for high-data-rate wireless networks, 
  such as data links between satellites and ground stations \cite{Toyoshima, OICETS, MonaRisa, NASAphotonreci}, ad hoc trunk link not bounded by fiber networks \cite{Military}, 
  and the ``last mile" link from the fiber backbone to the client premises \cite{urban}.
  
  The high directionality of laser beam can make FSO communications more secure than RF ones.  
  However, it has been shown in \cite{Agaskar, Puryear} that FSO communications can still suffer from optical tapping risks, 
  especially when the main lobe of laser beam is considerably wider than the receiver size, 
  which is the case for optical links between moving terminals, and also between satellites and ground stations.  
  To establish the secrecy of confidential data communications, symmetric key cryptography is often used with a preshared secret key or a key exchanged via public key cryptosystems.  
  These crypto-schemes are based on mathematical problems which are practically impossible to solve using current computer resources.  
  Its security is often referred to as computational security.  
  
  Recently, an approach based on physical layer security attracts much attention as an alternative mechanism.  
  This is based on an appropriate coding technique designed by considering physical properties of the channels, i.e., 
  the main channel between the sender (Alice) and the legitimate receiver (Bob), and the wiretapper channel from Alice to an eavesdropper (Eve).  
  This coding is particularly called the wiretap channel coding \cite{WynerWiretap, csiskor}, and realizes the two functions at the same time in the physical layer; the reliability for Bob and the secrecy against Eve.  
  The secrecy ensured by this paradigm is referred to as information theoretic security (ITS), which can be everlasting, 
  in the sense that it can be proved that Eve cannot obtain meaningful information even by unforeseen mathematical insights or by off-line attacks with future advanced computers.

  Studies so far on physical layer security in wireless channels and system architecture issues are nicely reviewed in \cite{BBbook}. 
  An information-theoretically secure key exchange protocol over quasi-static wireless channels was proposed with a near-optimal LDPC (low density parity check)-based reconciliation method over a wide range of signal-to-noise ratios (SNRs) \cite{Bloch}.  
  Physical layer security of FSO communications has been discussed in \cite{Wang}, proposing a secret key agreement over fading channels with reciprocity, and clarifying dominating factors on the secret key rate.  
  In \cite{LopezMartinez}, analysis was made on likely wiretap scenarios and influences to secure FSO communication performances, in terms of the outage probability of non-zero secrecy capacity.  
  Mostafa and Lampe studied physical layer security for indoor visible light communications \cite{visibleindoor}, 
  and showed that secrecy rates can be increased by utilizing Eve's channel state information (CSI) via null-steering, or by adding artificial noises when Eve's CSI is not available.  
  
  \begin{figure*}
    \centering
      \begin{overpic}[width=14cm, bb=0 0 996 419]{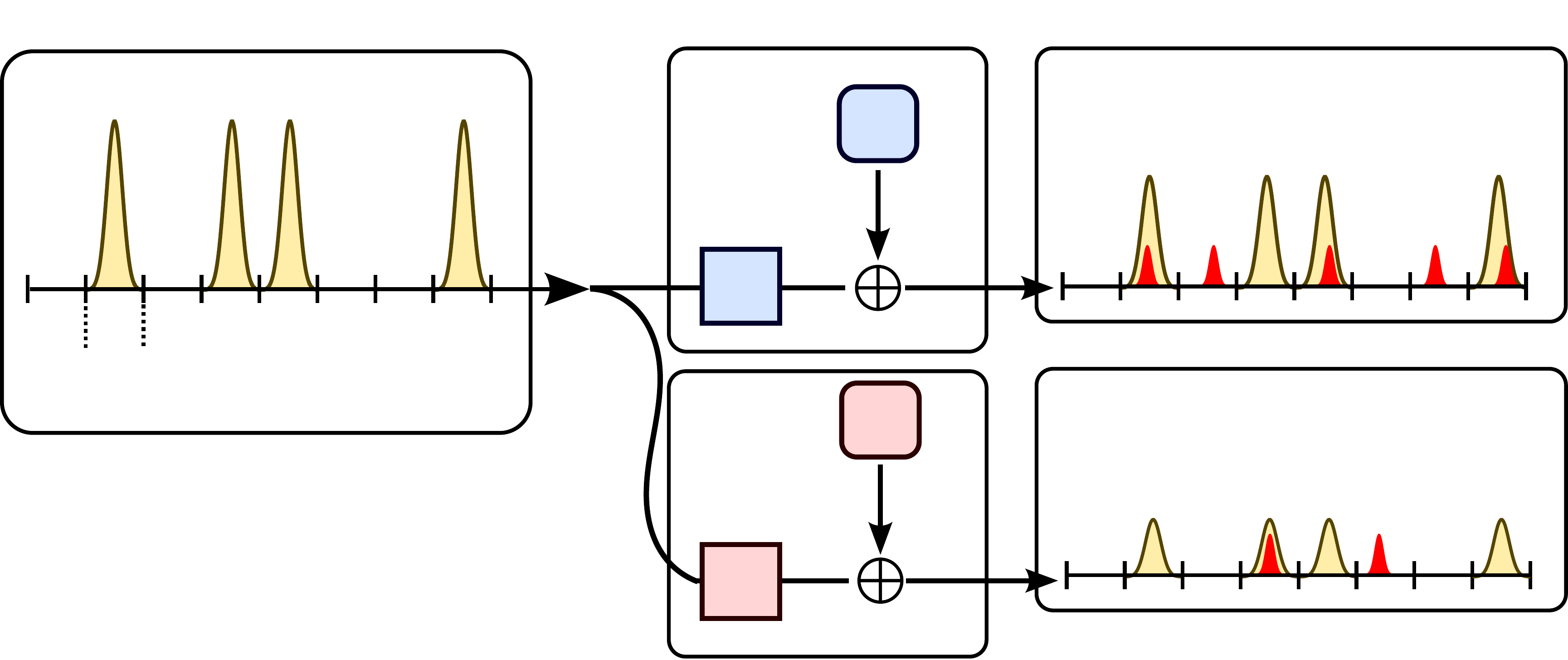}
        \put(55,158){\large{Alice}}
        \put(11,142.5){\large{$0$}}
        \put(26.5,142.5){\large{$1$}}
        \put(40.5,142.5){\large{$0$}}
        \put(56.5,142.5){\large{$1$}}
        \put(71.5,142.5){\large{$1$}}
        \put(85,142.5){\large{$0$}}
        \put(100,142.5){\large{$0$}}
        \put(115.5,142.5){\large{$1$}}
        \put(1,113){\large{$\Delta_{\mathrm{p}}$}}
        \put(17,116){\vector(1,0){9.9}}
        \put(41.5,116){\vector(-1,0){9.9}}
        \put(25,71){\large{$\Delta$}}
        \put(25,82){\vector(1,0){11}}
        \put(33,82){\vector(-11,0){11}}
        \put(95,74){\large{$n_A$}}
        \put(105,79){\vector(1,1.7){13}}
        \put(172,144){\large{$W_B$}}
        \put(183,94){\large{$\eta_y$}}
        \put(217,133){\large{$\lambda_y$}}
        \put(320,158){\large{Bob}}
        \put(308,134){\large{$n_B=\eta_y n_A$}}
        \put(305,135){\vector(-1,-1.4){13}}
        \put(351,118){\large{$\lambda_y\Delta$}}
        \put(351,121){\vector(-1,-1.6){15}}
        \put(172,61){\large{$W_E$}}
        \put(183,19){\large{$\eta_z$}}
        \put(217,58){\large{$\lambda_z$}}
        \put(320,1){\large{Eve}}
        \put(308,60){\large{$n_E=\eta_z n_A$}}
        \put(306,61){\vector(-1,-2.2){15}}
        \put(351,48){\large{$\lambda_z\Delta$}}
        \put(364,47){\vector(-1,-1.5){15}}
      \end{overpic}
    \caption{Wiretap channel based on on-off keying (OOK) modulation.  } \label{imple}
  \end{figure*}
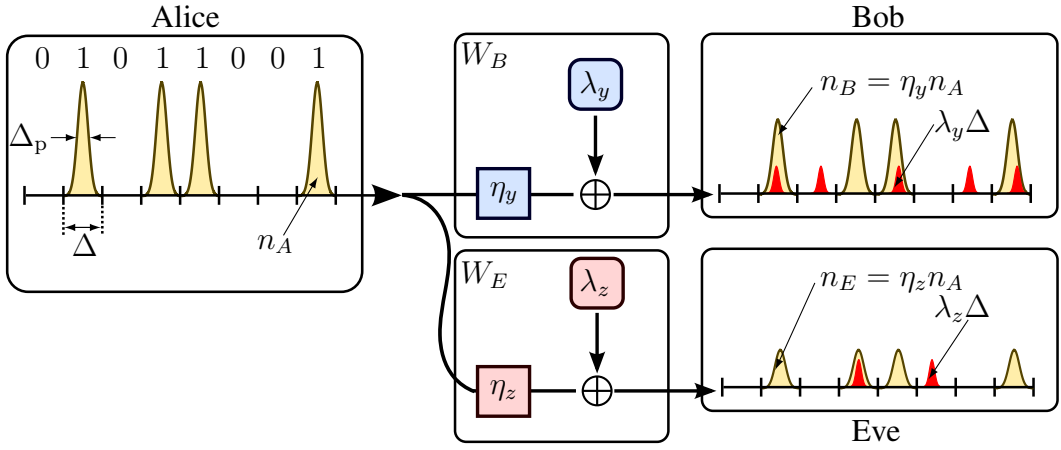
  
  An extreme example of physical layer security has been already realized in quantum key distribution (QKD) \cite{BB, E91, QKD}, 
  which has been extensively studied and now becomes practical in a metropolitan area fiber network \cite{SECOQC, TOKYOQKD}.  
  QKD ensures the unconditional security in the sense that Eve can have unlimited physical abilities and computational power.  
  For FSO channels which are basically line-of-sight (LoS) communications, however, this assumption is sometimes too much.  
  The LoS condition can naturally relax the assumption for Eve.  
  In fact, expected key rates of QKD in satellite-to-ground links are impractically poor if one insists on assuming that Eve can be everywhere in the universe and can do anything.  
  Instead, one should exploit more practical schemes which can attain higher key rate for LoS FSO channels under sensible assumptions case by case. 

  Design theory for wiretap channel coding should hopefully be able to evaluate the reliability for Bob and the secrecy against Eve.  
  Practically, the cost constraint at Alice's side, such as the power and bandwidth constraint, is an important factor to be cared.  
  In fact, transmission power should be carefully regulated so as not to increase wiretap risks.  
  Furthermore, the performances should eventually be characterized in finite length coding for practical use.  
  These issues have been partly dealt with in literatures \cite{HES, chou}, 
  but insights into unified theory and numerically expected performances have not been accumulated sufficiently yet, even in the idealistic setting of fading free channels.  

  In this paper, we study the optical wiretap channels with linear attenuation and background noises based on the on-off keying (OOK) modulation.  
  From the practical viewpoint, we impose the power constraint on Alice's available transmission power.  
  We numerically study the achievable secrecy rates and the secrecy capacity as a function of channel attenuation.  
  We compare them with the secure key rate for QKD to show how the performance can be increased by compromising the assumption on Eve.  
  According to the calculation, even if Eve can obtain $99\%$ as much power as Bob, 
  FSO links with ITS would be possible between geostationary earth orbit (GEO) satellites and ground stations with currently available technologies.  
  A functional meaning of auxiliary random variable originally introduced by Csisz\'ar and K\"orner \cite{csiskor} to establish the rate region of the general wiretap channel is clarified as a booster mechanism of the distance limit due to the auxiliary noises.  
  We then apply a recent theory on the error and secrecy exponents by some of authors \cite{HES} to finite length analysis on the optical wiretap channels.  
  We show how the code length to reach the given required levels of reliability and secrecy is estimated via the finite length analysis.  

  The paper is organized as follows.  
  In Section \ref{sec2}, we give the model and formulate the problems.  
  In Section \ref{sec3}, we present numerical results of an achievability rate (lower bound to the secrecy capacity) 
  and the structure of optimal parameters and power regulation.  
  Section \ref{sec4} includes analysis with the auxiliary random variable used at Alice.  
  Section \ref{sec5} describes the estimation of the necessary code length for the given required levels of performances via the finite length analysis.  
  The paper is concluded in Section \ref{conclu}. 
  
\section{Formulation of the model} \label{sec2}

  Throughout this paper, we consider a model of optical wiretap channel with linear attenuation and background noises based on on-off keying (OOK) modulation as shown in Fig. \ref{imple}.  
  This model consists of the main channel $W_B$ with which Alice transmits a confidential message to Bob
  and the wiretapper channel $W_E$ with which Eve attempts to observe the confidential message.  
  Bob and Eve receive the OOK signals by an on-off detector based on photon counting.  
  The main and wiretapper channels are characterized by two parameters: 
  the channel transmittances $\eta_y$ and $\eta_z$, and the dark count rates (DCR) $\lambda_y$ [counts/sec] (cps) and $\lambda_z$ [cps], respectively.  
  In this work, we dare to assume that the channels are fading free, in order to derive potentially achievable performances in good propagation conditions.  
  
  Alice is subject to the constraint with the maximum available transmission power of $P$ [W], 
  and transmits on- and off-signals encoding symbols ``1" and ``0" with probabilities $q$ and $1-q$, respectively. 
  The on-signal ``1" is conveyed by a laser pulse of width $\Delta_{\mathrm{p}}$ [s] and an average photon number $n_A$.  
  The off-signal ``0" is conveyed by the vacuum pulse.  
  Bob and Eve receive the attenuated pulses of the average photon numbers $n_B = \eta_y n_A$ and $n_E = \eta_z n_A$, respectively.  
  Detector efficiencies are renormalized into the channel transmittances.  
  In order to compare the fraction of power received by two parties, we introduce the relative transmittance $\eta_{zy} \equiv \eta_z/\eta_y$.  
  In the LoS scenario, $\eta_{zy} \le 1$ can be valid.  
  The detector time resolutions for Bob and Eve are finite, and assumed to be the same, $\Delta$ [s], for simplicity, and to be larger than the laser pulse width, i.e., $\Delta > \Delta_{\mathrm{p}}$.  
  This time resolution actually sets the maximum limit of repetition rate of optical pulses. 
  
  The above channel model should be regarded as a practical reduction of Poisson channel \cite{kabanov, wyner1}, which assumes an arbitrary short time resolution $\Delta \to 0$, 
  i.e., an infinite detector bandwidth, and 
  has been extensively studied in \cite{Wagner}, where the analytical formulas of the secrecy capacity were derived.  
  
  In the following, we mathematically formulate the model mentioned above.   

\subsection{Power constraint}
  Alice needs to optimize the input probability $q$ and the average photon number $n_A$ within the maximum available transmission power $P$.  
  In this work, we consider an optical channel at a center frequency $f_0=200$ THz (wavelength of $1.5$ $\mathsf{\mu}$m, which is eye safe and commonly used in optical fiber communications) with a certain bandwidth $B$ [Hz].  
  The pulsed laser of Alice is assumed to be Fourier-transform limited, i.e., $B \Delta_{\mathrm{p}}=1$.  
  The value of $B$ must be larger than the detector bandwidth $\Delta ^{-1}$.  
  For simplicity, an average photon number at each frequency, $\bar{n}(f)$, of the on-signal pulse is assumed to be the same value $n_A$ within the bandwidth $B$.  
  Thus the power per on-signal pulse is
    \begin{align}
      P_{\mathrm{p}} = \int^{\infty}_{-\infty} \bar{n}(f) h f df \simeq \int^{f_0+B/2}_{f_0-B/2} n_A hf df = \frac{n_A h f_0}{\Delta_{\mathrm{p}}},  
    \end{align}
  where $h$ is Planck's constant.  The total power of the OOK transmission is then
    \begin{align}
      P_{\mathrm{total}} = q \frac{\Delta_{\mathrm{p}}}{\Delta} P_{\mathrm{p}} = q \frac{n_A h f_0 }{\Delta},
      \end{align}
  which must be constrained by the maximum available power $P$.  
  Thus, we have the following power constraint: 
    \begin{align}
      q \frac{n_A h f_0}{\Delta} \leq P. \label{powerconst}
    \end{align}

\subsection{Channel matrices}
    \begin{figure}[tp]
      \centering
      \begin{picture}(170,110)(0,0)
        \thicklines
        \multiput(32.4,104.4)(2,-1.82){55}{\color{Salmon}{\line(1,-0.91){1}}}
        \multiput(32.4,71.6)(2,-0.62){55}{\color{Salmon}{\line(1,-0.31){1}}}
        \put(32.4,104.4){\color{red}{\line(1,-0.61){109.6}}}
        \put(32.4,71.6){\color{red}{\line(1,-0.61){109.6}}}
        \multiput(32.4,104.4)(2,-0.6){55}{\color{Gray}{\line(1,-0.3){1}}}
        \multiput(32.4,71.6)(2,0.6){55}{\color{Gray}{\line(1,0.3){1}}}
        \put(32.4,104.4){\line(1,0){109.6}}
        \put(32.4,71.6){\line(1,0){109.6}}
        \put(-2,102.1){\small{$x=0$}}
        \put(29.8,104.4){\circle{6}}
        \put(-2,69.4){\small{$x=1$}}
        \put(29.8,71.6){\circle{6}}
        \put(83,110){\small{$W_B$}}
        \put(152,102.1){\small{$y=0$}}
        \put(144.7,104.4){\circle{6}}
        \put(152,69.4){\small{$y=1$}}
        \put(144.7,71.6){\circle{6}}
        \put(152,34.2){\small{\textcolor{red}{$z=0$}}}
        \put(144.7,36.4){\color{red}{\circle{6}}}
        \put(152,1.5){\small{\textcolor{red}{$z=1$}}}
        \put(144.7,3.7){\color{red}{\circle{6}}}
        \put(83,25){\color{red}{\small{$W_E$}}}
        \put(120,80){\small{$a_y$}}
        \put(120,63){\small{$b_y$}}
        \put(120,27){\small{\textcolor{red}{$a_z$}}}
        \put(120,8){\small{\textcolor{red}{$b_z$}}}
      \end{picture}
      \caption{Channel diagram of wiretap channel.} \label{transitionprobability}
    \end{figure}
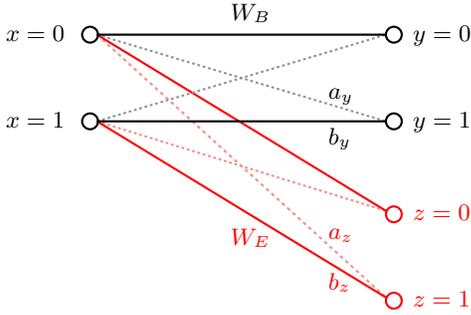
    
    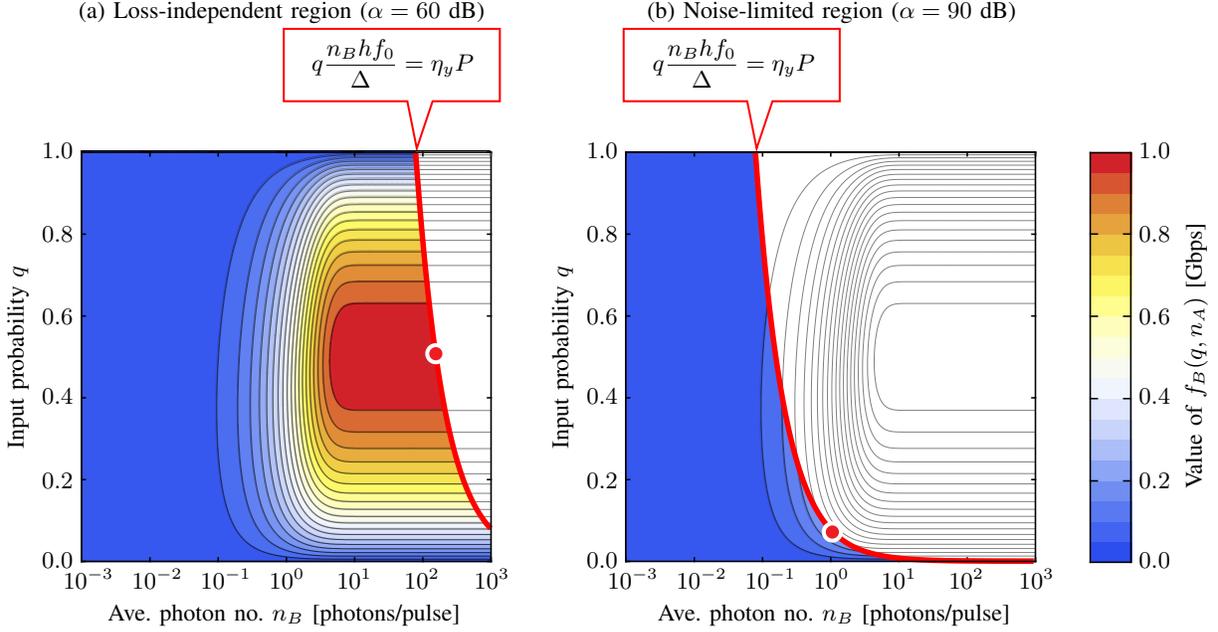
\begin{figure*}[tp]
      \centering
      \begin{overpic}[width=16cm,bb = 0 0 1242 664]{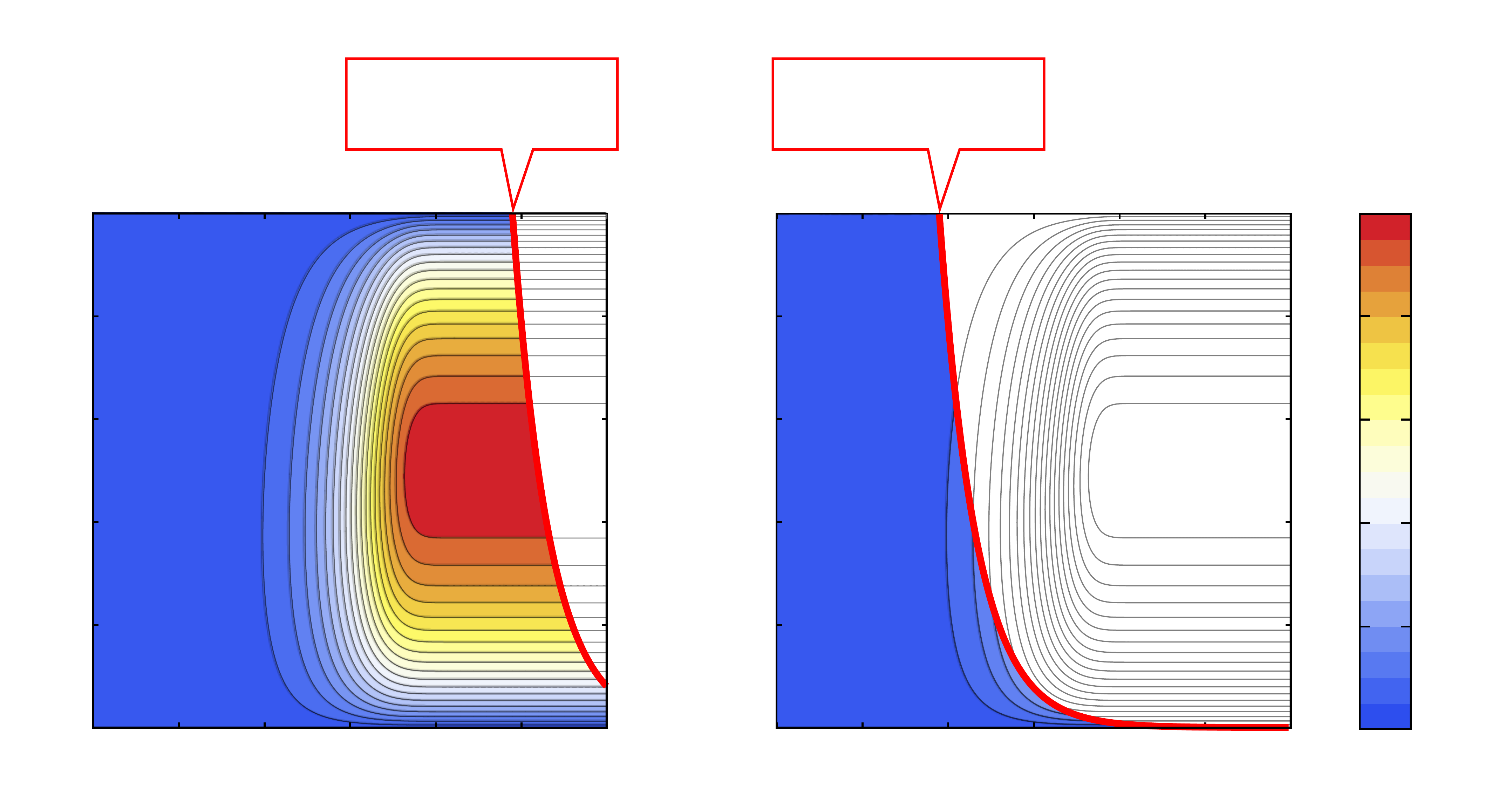}
        \put(38,2){\small{Ave. photon no. $n_B$ [photons/pulse]}}
        \put(21,14){\small{$10^{-3}$}}
        \put(47,14){\small{$10^{-2}$}}
        \put(73,14){\small{$10^{-1}$}}
        \put(99,14){\small{$10^{0}$}}
        \put(124.5,14){\small{$10^{1}$}}
        \put(150,14){\small{$10^{2}$}}
        \put(176.5,14){\small{$10^{3}$}}
        \put(0,67){\rotatebox{90}{\small{Input probability $q$}}}
        \put(13,22){\small{$0.0$}}
        \put(13,52){\small{$0.2$}}
        \put(13,84){\small{$0.4$}}
        \put(13,114){\small{$0.6$}}
        \put(13,145.5){\small{$0.8$}}
        \put(13,177){\small{$1.0$}}
        \put(115,210){\small{$\displaystyle q \frac{n_B h f_0}{\Delta} = \eta_y P$}}
        \put(27,230){\small{(a) Loss-independent region ($\alpha=60$ dB)}}
        \put(246,2){\small{Ave. photon no. $n_B$ [photons/pulse]}}
        \put(227,14){\small{$10^{-3}$}}
        \put(253,14){\small{$10^{-2}$}}
        \put(279,14){\small{$10^{-1}$}}
        \put(305,14){\small{$10^{0}$}}
        \put(330.5,14){\small{$10^{1}$}}
        \put(356,14){\small{$10^{2}$}}
        \put(382.5,14){\small{$10^{3}$}}
        \put(206,67){\rotatebox{90}{\small{Input probability $q$}}}
        \put(219,22){\small{$0.0$}}
        \put(219,52){\small{$0.2$}}
        \put(219,84){\small{$0.4$}}
        \put(219,114){\small{$0.6$}}
        \put(219,145.5){\small{$0.8$}}
        \put(219,177){\small{$1.0$}}
        \put(244,210){\small{$\displaystyle q \frac{n_B h f_0}{\Delta} = \eta_y P$}}
        \put(242,230){\small{(b) Noise-limited region ($\alpha=90$ dB)}}
        \put(428,22){\small{$0.0$}}
        \put(428,52){\small{$0.2$}}
        \put(428,84){\small{$0.4$}}
        \put(428,114){\small{$0.6$}}
        \put(428,145.5){\small{$0.8$}}
        \put(428,177){\small{$1.0$}}
        \put(445,52){\rotatebox{90}{\small{Value of $f_B(q,n_A)$ [Gbps]}}}
        \put(162,103){\color{White}{\circle*{8}}}
        \put(162,103){\color{Red}{\circle*{5}}}
        \put(312,35.5){\color{White}{\circle*{8}}}
        \put(312,35,5){\color{Red}{\circle*{5}}}
      \end{overpic}
      \caption{
      Contour plots of $f_B(q,n_A)$ as a function of input probability $q$ and average photon number $n_B = \eta_y n_A$ of the received pulse.  
      (a) The loss-independent region with $\alpha = -\log_{10} \eta_y = 60$ dB.  
      (b) The noise-limited region with $\alpha = 90$ dB.  
      The red circle and the white painted area in each plot represent the channel capacity $C$ and the non-allowed region due to the power constraint, respectively.  
      Parameter values: $P=10$ mW, $\lambda_y=10$ kcps, $\Delta=1$ ns.  } \label{zu301}
    \end{figure*}
    
  The symbols for Alice, Bob, and Eve are defined as $x$, $y$, and $z$, drawn from the binary random variables $X$, $Y$, and $Z$, respectively. 
  The on-off detectors at Bob and Eve discriminate the signals by the absence or presence of counts as ``0" or ``1".  
  Since the system is assumed to be stationary and memoryless, 
  the main channel $W_B$ illustrated in Fig. \ref{transitionprobability} can be fully described by the elementary channel with the channel matrix given as
    \begin{align}
      W_B(1|0) &= 1-e^{-\lambda_y \Delta} \equiv a_y, \non \\
      W_B(1|1) &= 1-e^{-(\eta_y n_A+\lambda_y \Delta)} \equiv b_y, \non
    \end{align}
  and
    \begin{align}
      W_B(0|0) &= e^{-\lambda_y \Delta} = 1-a_y, \non \\
      W_B(0|1) &= e^{-(\eta_y n_A+\lambda_y \Delta)} = 1-b_y, \non
    \end{align}
  Note that the DCR $\lambda_y$ is understood to include not only the dark counts of the detector but also the background noises in the main channel.  
  Similarly, the elements of the channel matrix of the wiretapper channel $W_E$ are given by
    \begin{align}
      W_E(1|0) &= 1-e^{-\lambda_z \Delta} \equiv a_z, \non \\
      W_E(1|1) &= 1-e^{-(\eta_z n_A+\lambda_z \Delta)} \equiv b_z, \non \\
      W_E(0|0) &= e^{-\lambda_z \Delta} = 1-a_z, \non \\
      W_E(0|1) &= e^{-(\eta_z n_A+\lambda_z \Delta)} = 1-b_z. \non
    \end{align}

\subsection{Channel capacity and secrecy rate}\label{sec23}
  In this subsection, we introduce necessary measures and formulas to evaluate the performance of our model.  
  In particular, starting with channel capacity, we provide the formula for achievable secrecy rate maximized over possible 
  transmission strategies without the auxiliary random variable $V$.
  The secrecy capacity is defined as the maximum achievable secrecy rate optimized also over the auxiliary random variable $V$ 
  in addition to the input variable $X$ \cite{csiskor}, because the additional randomness with $V$ can be helpful 
  for deceiving Eve especially when the wiretapper channel $W_E$ is not worse than the main channel $W_B$, and hence can improve the secrecy rate.  
  We will work on it later in Section \ref{sec4}.  

  Considering the standard channel coding without Eve, the maximum achievable rate of reliable transmission is called channel capacity, and is given by
    \begin{equation}
      C = \max_{P_X} I(X;Y), \label{defofchancap}
    \end{equation}
  where $I(X;Y)$ is the mutual information between the random variables $X$ and $Y$. 
  The maximization is taken over all possible input probability distribution $P_X$.  
  
  In this paper, we extend the above definition slightly so that not only the input probability $q$ but also the input signal intensity (the average photon number $n_A$) are simultaneously optimized under the power constraint (\ref{powerconst}).  
  Therefore, the channel $W_B$ is not a given fixed matrix but a 2-by-2 matrix variable through the parameter $n_A$ to be optimized.  
  The channel capacity is then defined as 
    \begin{equation}
      C \equiv \max_{q,n_A} f_B(q,n_A), \label{channelcapacityOnOff}
    \end{equation}
  where
    \begin{align}
      f_B(q,n_A) \equiv &h_2 ((1-q)a_y + q (1-b_y)) \non \\
                        & \quad - (1-q) h_2 (a_y) - q h_2 (b_y), \label{mimainPPM}
    \end{align}
  with the binary entropy function defined as
    \begin{equation}
      h_2 (q) \equiv - q \log_2 q - (1-q) \log_2 (1-q).
    \end{equation}

  In the wiretap channel coding, we concern the asymptotically maximum achievable secrecy rate of reliable transmission to Bob while ensuring the ITS against Eve, 
  which is defined in the form as \cite{WynerWiretap}
    \begin{equation}
      R_\mathrm{S} = \max_{P_X} \left[I(X;Y) - I(X;Z)\right]. \label{defofseccap}
    \end{equation}

  To have a positive value of $R_\mathrm{S}$, the relation $I(X;Y) \ge I(X;Z)$ should hold for any $X$,  
  which means that the main channel $W_B$ is better than the wiretapper channel $W_E$ regardless of the input strategy.  
  If this is the case, the wiretap channel is said to be more capable
  and the above quantity coincides with the secrecy capacity, which will be mentioned later in Section \ref{sec4}. 
  In this paper, we deal with general cases, not necessarily being more capable, by assuming that the wiretapper channel $W_E$ is not worse. 
  It depends on $\eta_y, \lambda_y, \eta_z$ and $\lambda_z$ whether the wiretap channel is more capable or less capable.  
  Now the similar extension for the simultaneous optimization of $q$ and $n_A$ is made as 
    \begin{equation}
      R_\mathrm{S} \equiv \max_{q,n_A} f_{BE}(q,n_A), \label{secrecyX}
    \end{equation}
  where  
    \begin{equation}
      f_{BE}(q,n_A) \equiv f_B(q,n_A) - f_E(q,n_A),
    \end{equation}
  and
    \begin{align}
      f_E(q,n_A) \equiv & h_2((1-q)a_z + q (1-b_z)) \non \\
                        & \quad - (1-q)h_2(a_z) - qh_2(b_z).
    \end{align}
  
\section{Numerical results of channel capacity and secrecy rate} \label{sec3}
  It is generally difficult to derive a closed form expression for the channel capacity (\ref{channelcapacityOnOff}) and the secrecy rate (\ref{secrecyX}) except for simple channels such as a binary symmetric channel.  
  Hence, we carry out the numerical optimization in order to obtain these quantities.  
  Throughout this section, we adopt a set of parameters as follows: $P=10$ mW, $\lambda_y=10$ kcps, $\lambda_z=1$ cps, $\Delta=1$ ns,  
  where the value of the time resolution $\Delta$ corresponds to the maximum possible pulse repetition rate of $1$ GHz.  
  Note that the above parameters represent the case where Alice and Bob have the transmitter and the detector which will be available at the current level of technology, respectively, whereas Eve may have a much less noisy detector.  
  In this case, the wiretap channel is not more capable for all possible values of $n_A$.
  
\subsection{Channel capacity}\label{sec32}
  In this subsection, we present basic results of the channel capacity when there is nothing to do with the wiretapper channel, 
  discuss important features in our model, and prepare ourselves for the main analysis on the secrecy rate.  

  Fig. \ref{zu301} shows contour plots of the mutual information $f_B(q,n_A)$ as a function of input probability $q$ and average photon number $n_B = \eta_y n_A$ of the received pulse.  
  The calculations are demonstrated for two typical cases, (a) for a sufficiently small attenuation $\alpha$ (short distance transmission) where an attenuation $\alpha$ is defined by $\alpha = - \log_{10} \eta_y$, 
  and (b) for a larger attenuation $\alpha$ (long distance transmission).  
  From this figure, we can know how the channel capacity and the optimal $q$ and $n_B$ (and hence $n_A$) are determined as the attenuation $\alpha$ varies.  
  The power constraint translated in terms of received power at Bob is represented by the left lower region below the boundary (red solid line), which is referred to as the allowed region.  
  The channel capacity $C$, indicated by the red circle, can be found on this boundary line.  
  The right upper region is not allowed by the power constraint, referred to as the non-allowed region.  

  In Fig. \ref{zu301}(a), the power constraint border (red line) crosses the plateau of the maximum value of $f_B(q,n_A)$.  
  As the attenuation $\alpha$ increases (the amount of the received power $\eta_y P$ decreases), the non-allowed region (right-upper area) extends to the left-lower side.  
  Unless the power constraint border gets out of the plateau of the maximum of $f_B(q,n_A)$, the value of the channel capacity remains the same value, independent of $\alpha$.  
  In this region, Alice's power is sufficient enough to transmit the signals such that Bob's detector can well discriminate them, not limited by the noises.  
  We refer to the region as the loss-independent region. 

  When the power constraint border has once gotten out of the plateau of the maximum of $f_B(q,n_A)$ as depicted in Fig. \ref{zu301}(b), the channel capacity starts to decrease.  
  One can see the optimal $q$ should also decrease.  
  This means that Alice had better to send the on-signal less frequently to be able to make the on-signal as bright as possible under the power constraint so that Bob's detector can discriminate it from the noise background with high SNR.  
  We refer to the region as the noise-limited region.   
    \begin{figure}[tp]
      \centering
      \begin{overpic}[width=8cm,bb = 0 0 485 471]{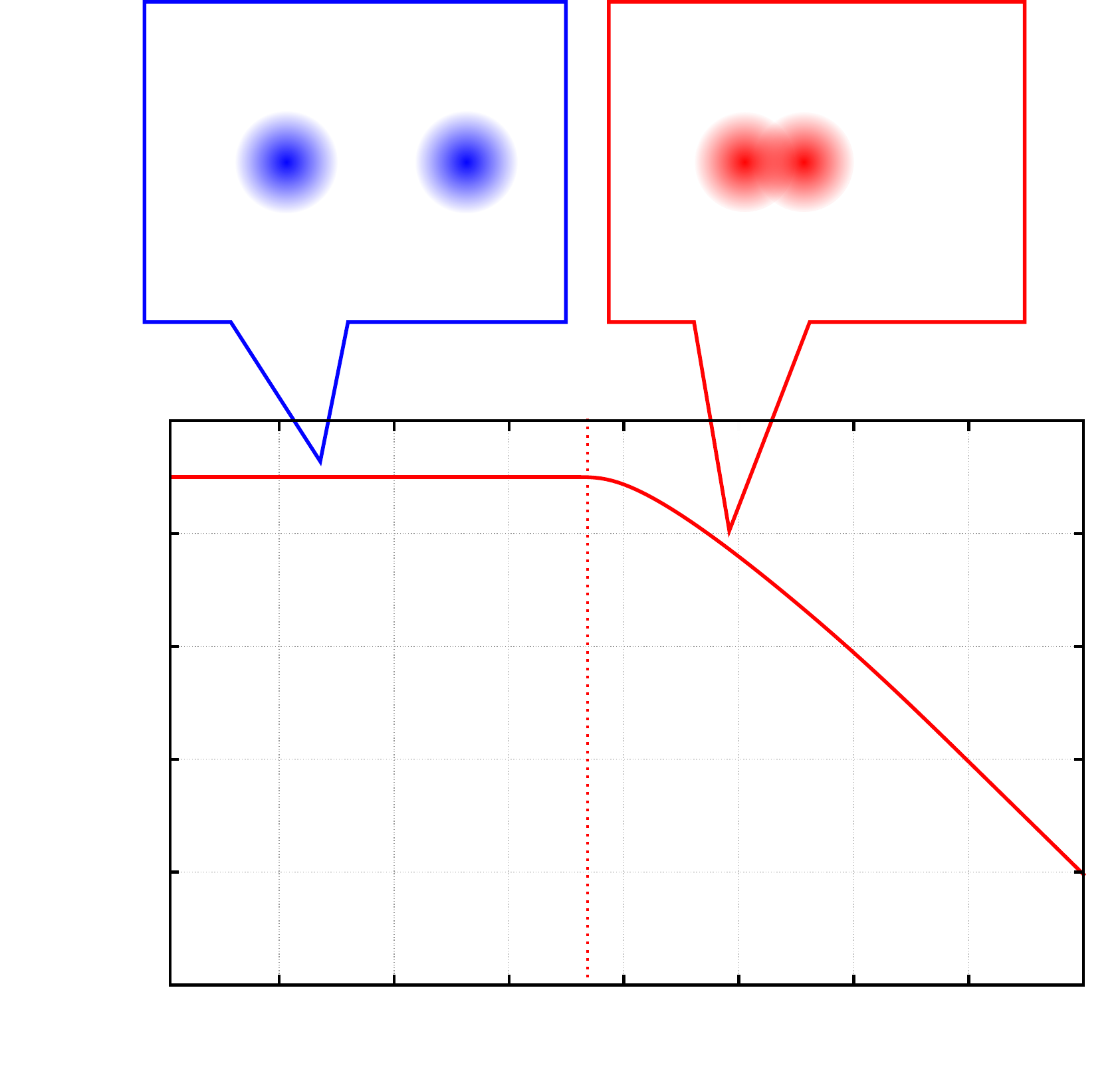}
        \put(91,0){\small{Attenuation $\alpha$ [dB]}}
        \put(32.2,12){\small{$0$}}
        \put(75.4,12){\small{$40$}}
        \put(122.3,12){\small{$80$}}
        \put(167.3,12){\small{$120$}}
        \put(214.3,12){\small{$160$}}
        \put(-2,25){\rotatebox{90}{\small{Channel capacity $C$ [bps]}}}
        \put(29.5,19){\small{$1$}}
        \put(18.5,41){\small{$100$}}
        \put(18.5,63.5){\small{$10$k}}
        \put(21,86){\small{$1$M}}
        \put(11,110){\small{$100$M}}
        \put(16,132.5){\small{$10$G}}
        \put(66,37){\small{region}}
        \put(42.5,47){\small{Loss-independent}}
        \put(37,33.5){\vector(1,0){82}}
        \put(117,33.5){\vector(-1,0){82}}
        \put(133,37){\small{Noise-limited region}}
        \put(122.5,33.5){\vector(1,0){97}}
        \put(217,33.5){\vector(-1,0){97}}
        \put(60,213){\small{$Q$}}
        \put(105,190){\small{$I$}}
        \put(75,200){\small{$y=1$}}
        \put(35,200){\small{$y=0$}}
        \put(153,213){\small{$Q$}}
        \put(200,190){\small{$I$}}
        \put(165,200){\small{$y=1$}}
        \put(128,200){\small{$y=0$}}
        \thicklines
        \put(58.2,217){\line(0,-1){57}}
        \put(48,188.1){\line(1,0){57}}
        \put(151.3,217){\line(0,-1){57}}
        \put(141,188.1){\line(1,0){57}}
        \thinlines
        \put(65.4,123.7){\color{Blue}{\circle*{6}}}
        \put(65.4,123.7){\color{White}{\circle*{4}}}
        \put(148.1,109.6){\color{Red}{\circle*{6}}}
        \put(148.1,109.6){\color{White}{\circle*{4}}}
      \end{overpic}
      \caption{Channel capacity $C$ as a function of attenuation $\alpha = - \log_{10} \eta_y$.  
      The upper insets are the intensity-quadrature constellations for received signals in the loss-independent (left) region and the noise-limited region (right).  
      Parameters: $P=10$ mW, $\lambda_y=10$ kcps, $\Delta=1$ ns.  } \label{channelcapacity}
    \end{figure}
    \begin{figure}[tp]
      \centering
      \begin{overpic}[width=8cm, bb = 0 0 480 329]{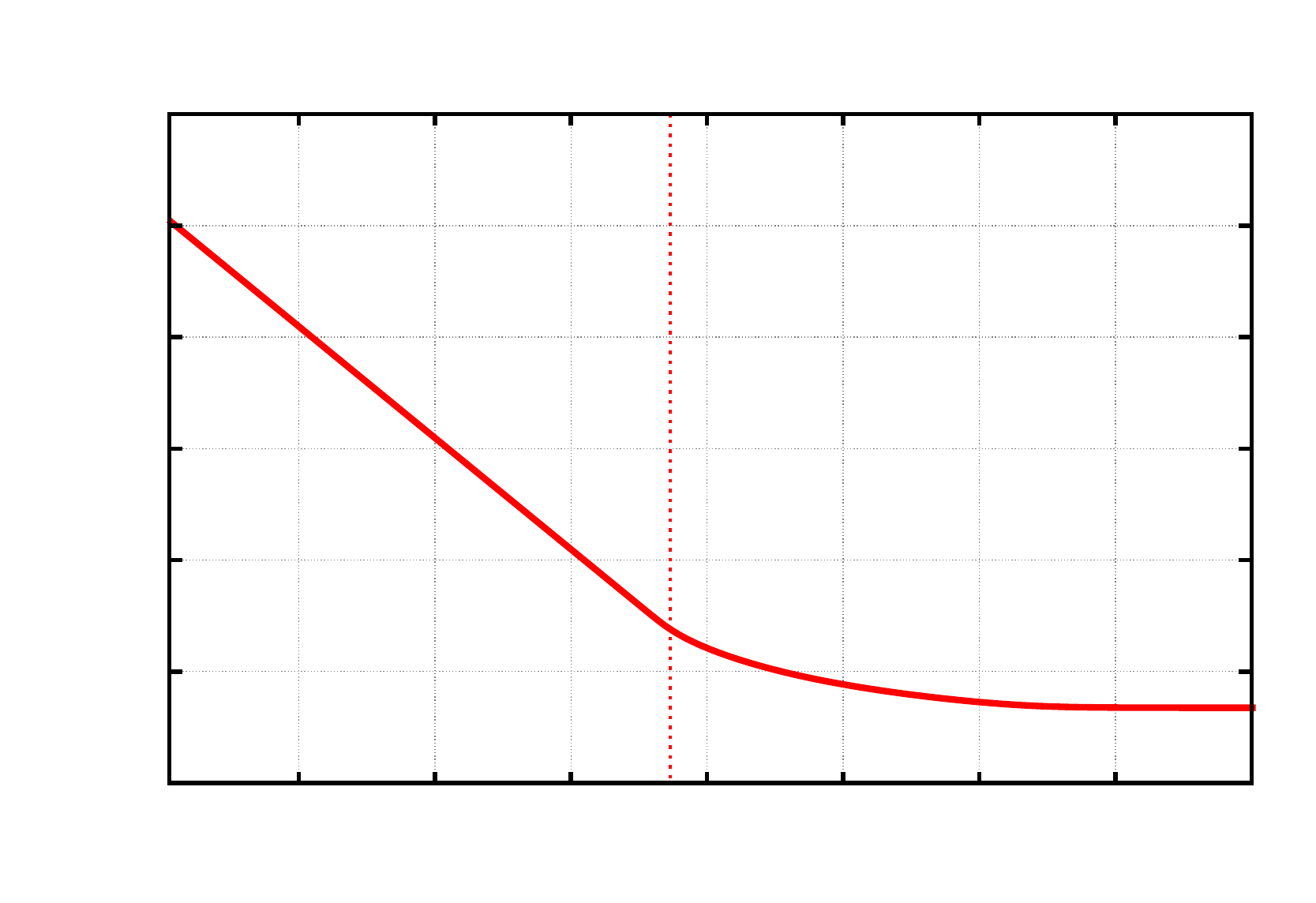}
        \put(85,3){\small{Attenuation $\alpha$} [dB]}
        \put(26,12){\small{$0$}}
        \put(69.5,12){\small{$40$}}
        \put(116,12){\small{$80$}}
        \put(160.5,12){\small{$120$}}
        \put(207,12){\small{$160$}}
        \put(8,17.5){\small{$10^{-2}$}}
        \put(24,37){\small{$1$}}
        \put(14,56){\small{$10^2$}}
        \put(14,75){\small{$10^4$}}
        \put(14,94){\small{$10^6$}}
        \put(14,113){\small{$10^8$}}
        \put(10,132){\small{$10^{10}$}}
        \put(0,18){\rotatebox{90}{\footnotesize{Ave. photon no. $n^{\ast}_B$ [photons/pulse]}}}
        \put(40,117){\small{Loss-independent}}
        \put(60,107){\small{region}}
        \put(32,125){\vector(1,0){82}}
        \put(111,125){\vector(-1,0){82}}
        \put(127,117){\small{Noise-limited region}}
        \put(118,125){\vector(1,0){96}}
        \put(211,125){\vector(-1,0){96}}
      \end{overpic}
      \caption{Optimal average photon number $n^{\ast}_B$ for the channel capacity $C$ in Fig. \ref{channelcapacity}.} \label{channelcapacitynB}
    \end{figure}
    \begin{figure}[tp]
      \centering
      \begin{overpic}[width=8cm, bb = 0 0 480 329]{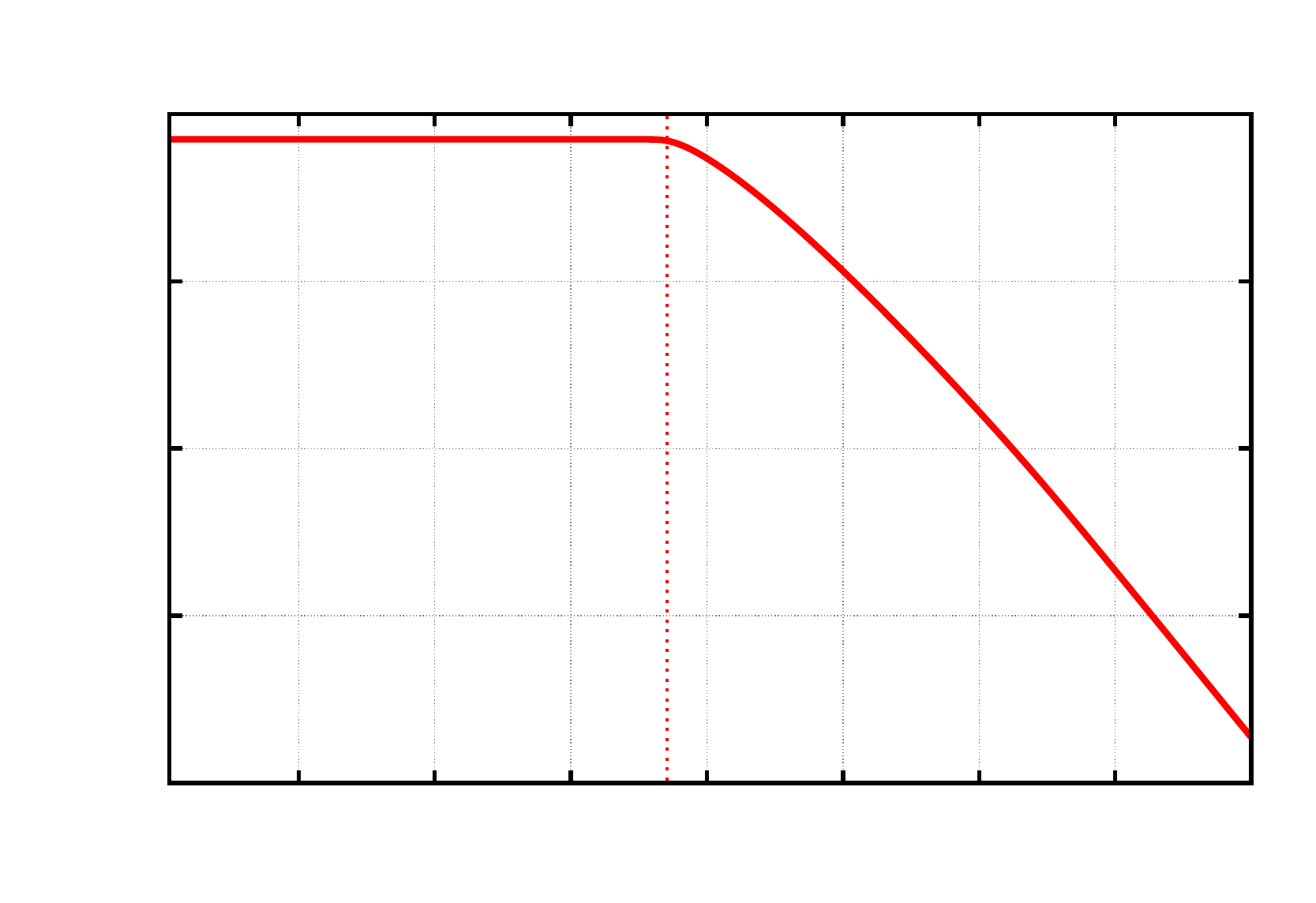}
        \put(85,3){\small{Attenuation $\alpha$} [dB]}
        \put(26,12){\small{$0$}}
        \put(69.5,12){\small{$40$}}
        \put(116,12){\small{$80$}}
        \put(160.5,12){\small{$120$}}
        \put(207,12){\small{$160$}}
        \put(8,17.5){\small{$10^{-8}$}}
        \put(8,46.5){\small{$10^{-6}$}}
        \put(8,75){\small{$10^{-4}$}}
        \put(8,103.5){\small{$10^{-2}$}}
        \put(24,132){\small{$1$}}
        \put(-2,42){\small{\rotatebox{90}{Input probability $q^{\ast}$}}}
        \put(40,37){\small{Loss-independent}}
        \put(60,28){\small{region}}
        \put(32,25){\vector(1,0){82}}
        \put(111.5,25){\vector(-1,0){82}}
        \put(127,28){\small{Noise-limited region}}
        \put(118,25){\vector(1,0){96}}
        \put(211,25){\vector(-1,0){96}}
      \end{overpic}
      \caption{Optimal input probability $q^{\ast}$ the channel capacity $C$ in Fig. \ref{channelcapacity}. } \label{channelcapacityq}
    \end{figure}
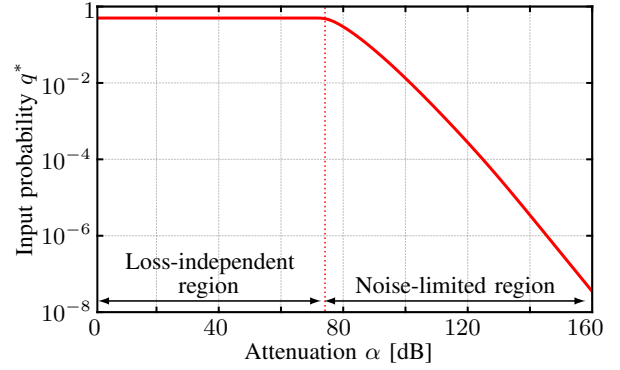

  Such behaviors can be explicitly seen in Fig. \ref{channelcapacity}, by the channel capacity $C$ as a function of attenuation $\alpha$.  
  The optimal parameters $n_B^{\ast}$ and $q^{\ast}$ are shown in Figs. \ref{channelcapacitynB} and \ref{channelcapacityq}, respectively.  
  In the loss-independent region, although $n_B^{\ast}$ decreases as $\alpha$ increases, Bob can still have a sufficiently high SNR, hence the capacity is unchanged.  
  The $q^{\ast}$ is about $0.5$.  
  In the noise-limited region, $n_B^{\ast}$ stays at a level of around $1$ photon/pulse so that the SNR for the received signals is not further degraded 
  (keeping the distance between the on- and off-signals in the I-Q constellation diagram the same order as the noise distribution), 
  while $q^{\ast}$ should decrease as $\alpha$ increases so that the power constraint is satisfied.  
  The channel capacity decreases as $\alpha$ increases, according roughly to $q^{\ast}$.  
  
\subsection{Secrecy rate}\label{sec33}
    \begin{figure*}[tp]
    \centering
      \begin{overpic}[width=16cm,bb = 0 0 1242 664]{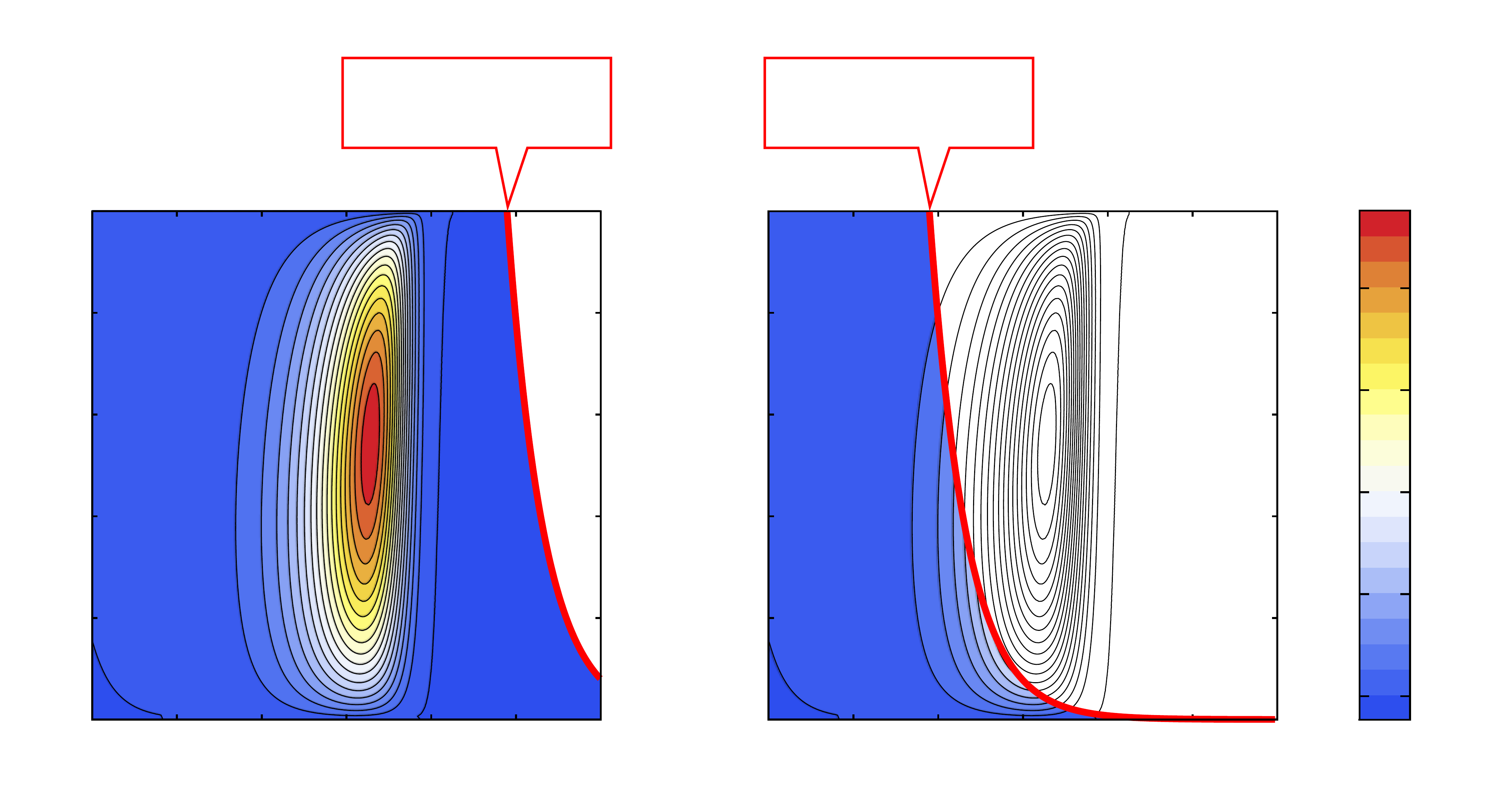}
        \put(38,2){\small{Ave. photon no. $n_B$ [photons/pulse]}}
        \put(21,14){\small{$10^{-3}$}}
        \put(47,14){\small{$10^{-2}$}}
        \put(73,14){\small{$10^{-1}$}}
        \put(99,14){\small{$10^{0}$}}
        \put(124.5,14){\small{$10^{1}$}}
        \put(150,14){\small{$10^{2}$}}
        \put(176.5,14){\small{$10^{3}$}}
        \put(0,67){\rotatebox{90}{\small{Input probability $q$}}}
        \put(13,22){\small{$0.0$}}
        \put(13,52){\small{$0.2$}}
        \put(13,84){\small{$0.4$}}
        \put(13,114){\small{$0.6$}}
        \put(13,145.5){\small{$0.8$}}
        \put(13,177){\small{$1.0$}}
        \put(115,210){\small{$\displaystyle q \frac{n_B h f_0}{\Delta} = \eta_y P$}}
        \put(27,232){\small{(a) Loss-independent region ($\alpha=60$ dB)}}
        \put(246,2){\small{Ave. photon no. $n_B$ [photons/pulse]}}
        \put(227,14){\small{$10^{-3}$}}
        \put(253,14){\small{$10^{-2}$}}
        \put(279,14){\small{$10^{-1}$}}
        \put(305,14){\small{$10^{0}$}}
        \put(330.5,14){\small{$10^{1}$}}
        \put(356,14){\small{$10^{2}$}}
        \put(382.5,14){\small{$10^{3}$}}
        \put(206,67){\rotatebox{90}{\small{Input probability $q$}}}
        \put(219,22){\small{$0.0$}}
        \put(219,52){\small{$0.2$}}
        \put(219,84){\small{$0.4$}}
        \put(219,114){\small{$0.6$}}
        \put(219,145.5){\small{$0.8$}}
        \put(219,177){\small{$1.0$}}
        \put(244,210){\small{$\displaystyle q \frac{n_B h f_0}{\Delta} = \eta_y P$}}
        \put(242,232){\small{(b) Noise-limited region ($\alpha=90$ dB)}}
        \put(432,28.5){\small{$0$}}
        \put(432,60){\small{$10$}}
        \put(432,91.5){\small{$20$}}
        \put(432,123){\small{$30$}}
        \put(432,153){\small{$40$}}
        \put(445,52){\rotatebox{90}{\small{Value of $f_{BE}(q,n_A)$ [Mbps]}}}
        \put(112.5,108){\color{White}{\circle*{8}}}
        \put(112.5,108){\color{Red}{\circle*{5}}}
        \put(305,45){\color{White}{\circle*{8}}}
        \put(305,45){\color{Red}{\circle*{5}}}	
      \end{overpic}
    \caption{
    Contour plots of $f_{BE}(q,n_A)$ as a function of input probability $q$ and average photon number $n_B$ of the received pulse and input probability $q$.  
    (a) The loss-independent region with $\alpha =-\log_{10} \eta_y = 60$ dB.  
    (b) The noise-limited region with $\alpha = 90$ dB. 
    The red circle and the white painted area denote the secrecy rate $R_\mathrm{S}$ and the non-allowed region due to the power constraint, respectively.  
    Parameters: $P=10$ mW, $\eta_{zy}=0.95$, $\lambda_y=10$ kcps, $\lambda_z=1$ cps, $\Delta=1$ ns. } \label{zu30}
  \end{figure*}
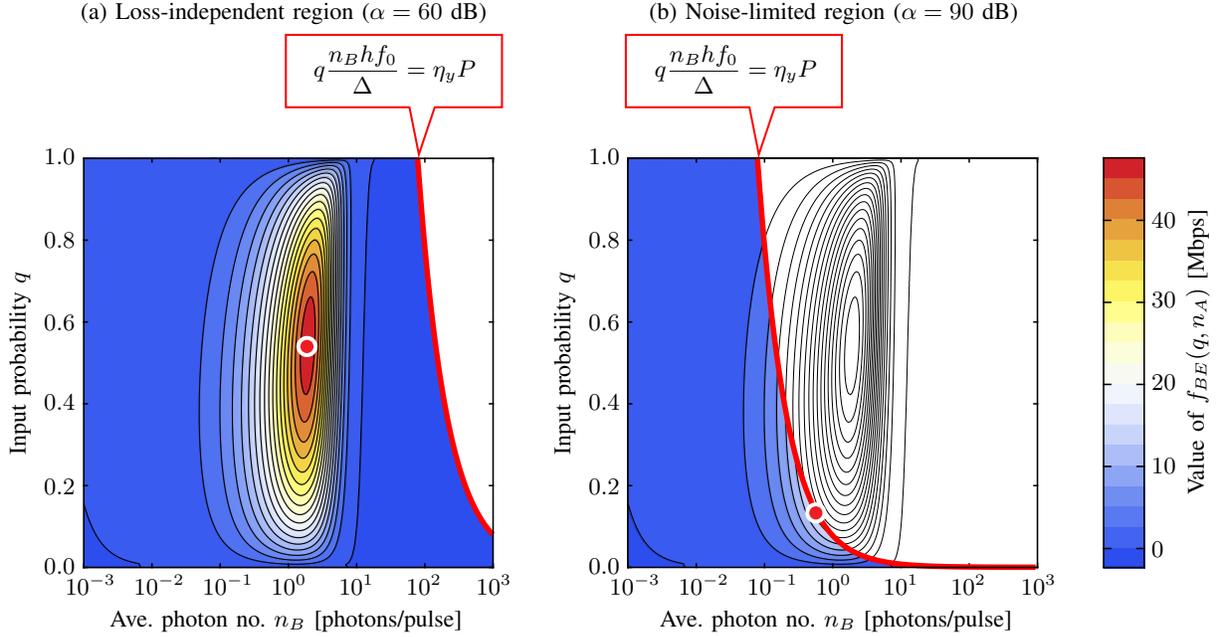
  In this subsection, based on the analysis carried out in the previous subsection, we move onto the main analysis on the secrecy rate $R_{\mathrm{S}}$, 
  and discuss the optimal strategy. 

  Fig. \ref{zu30} shows contour plots of $f_{BE}(q,n_A)$ as a function of input probability $q$ and average photon number $n_B$ of the received pulse.  
  Contrary to $f_B(q,n_A)$ shown in Fig. \ref{zu301}, the function $f_{BE}(q,n_A)$ sharply decreases at large $n_B$,  
  which is intuitively understood that the bright pulse increases the information leakage against Eve.  
  Moreover, the value of $f_{BE}(q,n_A)$ can be negative, because Bob's detector is much more noisy than Eve's one, and hence the wiretap channel is not more capable.  

  Fig. \ref{zu30}(a) is for the loss-independent region.  
  The maximum of $f_{BE}(q,n_A)$ (red circle) is located inside the allowed region.  
  Unless the power constraint border (red solid line) passes over this maximum to the left-lower side, the secrecy rate $R_\mathrm{S}$ can be realized at this maximum.  
  Thus, the optimal parameters $(q^{\ast},n_A^{\ast})$ satisfy the strict inequality as
    \begin{align}
      q^{\ast} \frac{n^{\ast}_A h f_0 }{\Delta} < P, \label{losslessconst}
    \end{align}
  indicating that Alice should not use the available power fully but regulate the transmission power properly 
  so as to prevent the confidential information from leaking against Eve.  

  Similarly to the channel capacity, the secrecy rate begins to decrease when the power constraint border line has once passed over the maximum of $f_{BE}(q,n_A)$ as shown in Fig. \ref{zu30}(b).  
  In this region, the secrecy rate is located on this border such that Alice should use all the available power to retain the necessary SNR.  
  Thus, the optimal parameters $(q^{\ast},n_A^{\ast})$ satisfy the power constraint with holding equality as
    \begin{align}
      q^{\ast} \frac{n^{\ast}_A h f_0 }{\Delta} = P. \label{equality}
    \end{align}
  
    \begin{figure}[tp]
      \centering
      \begin{overpic}[width=8cm,bb = 0 0 485 471]{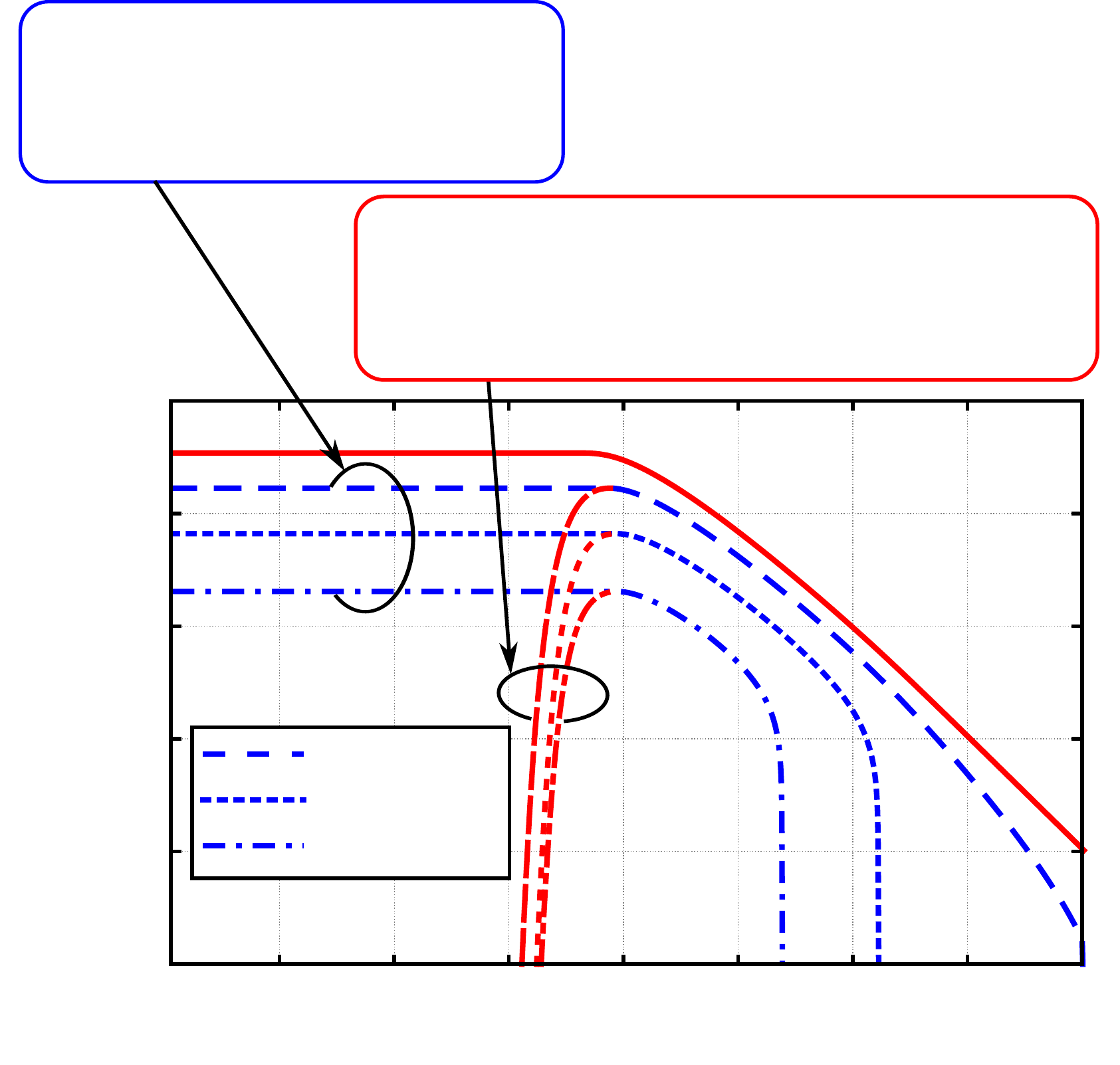}
        \put(91,0){\small{Attenuation $\alpha$ [dB]}}
        \put(32.2,12){\small{$0$}}
        \put(75.4,12){\small{$40$}}
        \put(122.3,12){\small{$80$}}
        \put(167.3,12){\small{$120$}}
        \put(214.3,12){\small{$160$}}
        \put(0,33){\rotatebox{90}{\small{Secrecy rate $R_\mathrm{S}$ [bps]}}}
        \put(29.5,19){\small{$1$}}
        \put(18.5,41){\small{$100$}}
        \put(18.5,63.5){\small{$10$k}}
        \put(21,86){\small{$1$M}}
        \put(11,110){\small{$100$M}}
        \put(16,132.5){\small{$10$G}}
        \put(6,207){\small{Power regulation is made}}
        \put(33,190){\small{$\displaystyle q^{\ast} \frac{n^{\ast}_A h f_0}{\Delta} < P$}}
        \put(76,167){\small{The available power is fully used as}}
        \put(123,150){\small{$\displaystyle q^{\ast} \frac{n^{\ast}_A h f_0}{\Delta} = P$}}
        \put(64,61){\footnotesize{$\eta_{zy}\!=0.5$}}
        \put(64,52){\footnotesize{$\eta_{zy}\!=0.9$}}
        \put(64,42.4){\footnotesize{$\eta_{zy}\!=0.99$}}
        \put(130,128){\rotatebox{-40}{\small{Channel capacity ($\eta_{zy}=0$)}}}
      \end{overpic}
      \caption{Secrecy rate $R_\mathrm{S}$ as a function of attenuation $\alpha = -\log_{10} \eta_y$ with various relative transmittances $\eta_{zy}$.    
      Also shown for comparison are the cases where the available power at Alice is used up.  
      Parameters: $P=10$ mW, $\lambda_y=10$ kcps, $\lambda_z=1$ cps, $\Delta=1$ ns.  } \label{zu35}
    \end{figure}
    \begin{figure}[tp]
      \centering
      \begin{overpic}[width=8cm,bb=0 0 480 329]{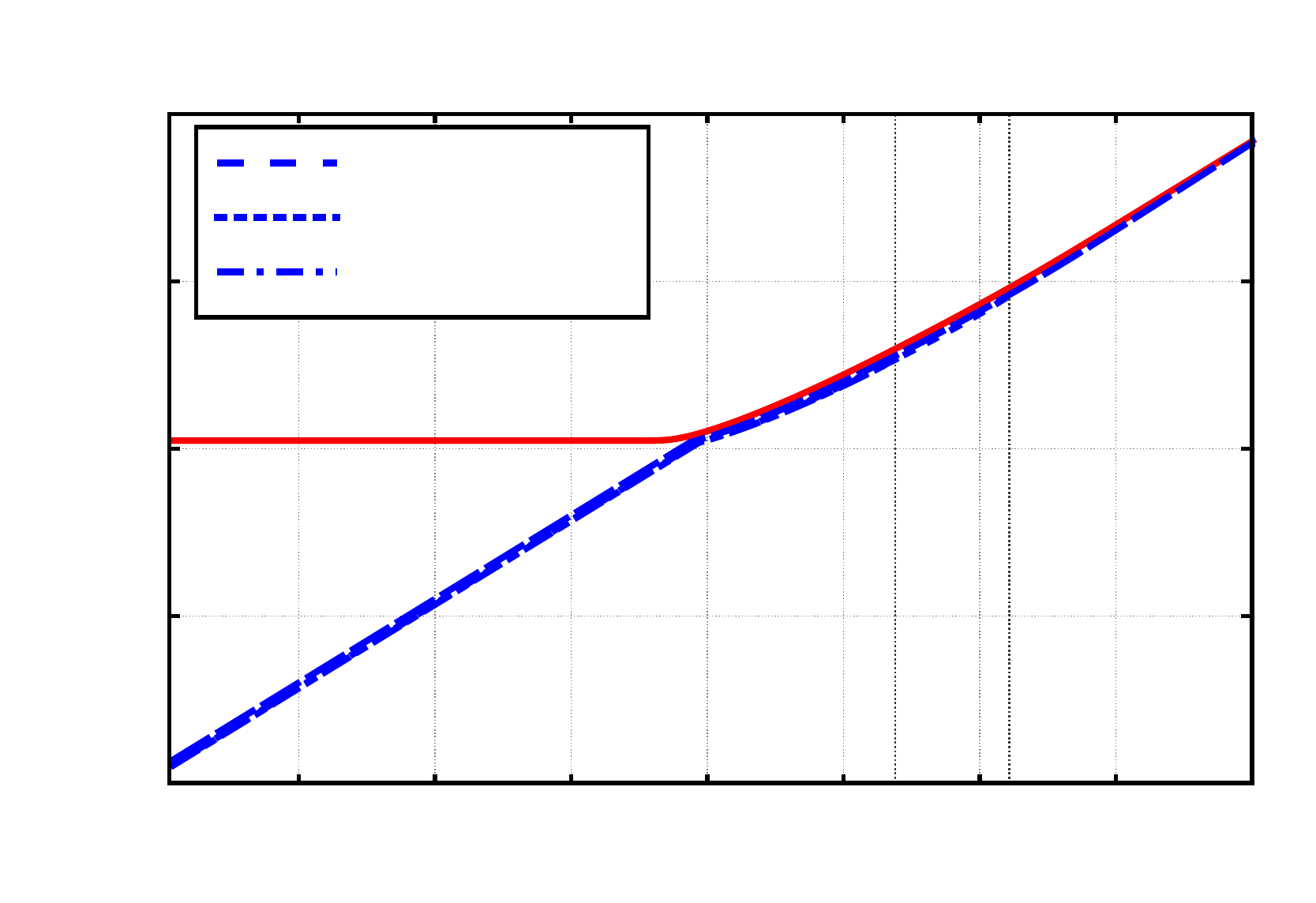}
        \put(35,83.5){\footnotesize{Without Eve ($\eta_{zy}=0$)}}
        \put(64,125.3){\small{$\eta_{zy}=0.5$}}
        \put(64,116){\small{$\eta_{zy}=0.9$}}
        \put(64,106.5){\small{$\eta_{zy}=0.99$}}
        \put(154.5,158){\small{$124.4$ dB}}
        \put(174.8,156){\vector(0,-1){19}}
        \put(126.5,148){\small{$107.6$ dB}}
        \put(154.9,147){\vector(0,-1){10}}
        \put(90,3){\small{Attenuation $\alpha$} [dB]}
        \put(26.5,12){\small{$0$}}
        \put(70.5,12){\small{$40$}}
        \put(117.5,12){\small{$80$}}
        \put(162,12){\small{$120$}}
        \put(209.5,12){\small{$160$}}
        \put(22,17.5){\small{$1$}}
        \put(14,46.5){\small{$10^4$}}
        \put(14,75){\small{$10^8$}}
        \put(10,103.5){\small{$10^{12}$}}
        \put(10,132){\small{$10^{16}$}}
        \put(0,18){\rotatebox{90}{\footnotesize{Ave. photon no. $n^{\ast}_A$ [photons/pulse]}}}
      \end{overpic}
      \caption{Optimal average photon number $n^{\ast}_A$ of the input pulse for the secrecy rate $R_\mathrm{S}$ in Fig. \ref{zu35}.  
      Since the secrecy rate decreases to $0$, the curves for $\eta_{zy} = 0.9$ and $\eta_{zy} = 0.99$ are shown up to $\alpha = 124.4$ dB and $107.6$ dB, respectively.  
      The solid line denotes the parameter for the case without Eve ($\eta_{zy}=0$) which leads to the channel capacity $C$.} \label{zu32}
    \end{figure}
    \begin{figure}[tp]
      \centering
      \begin{overpic}[width=8cm,bb=0 0 480 329]{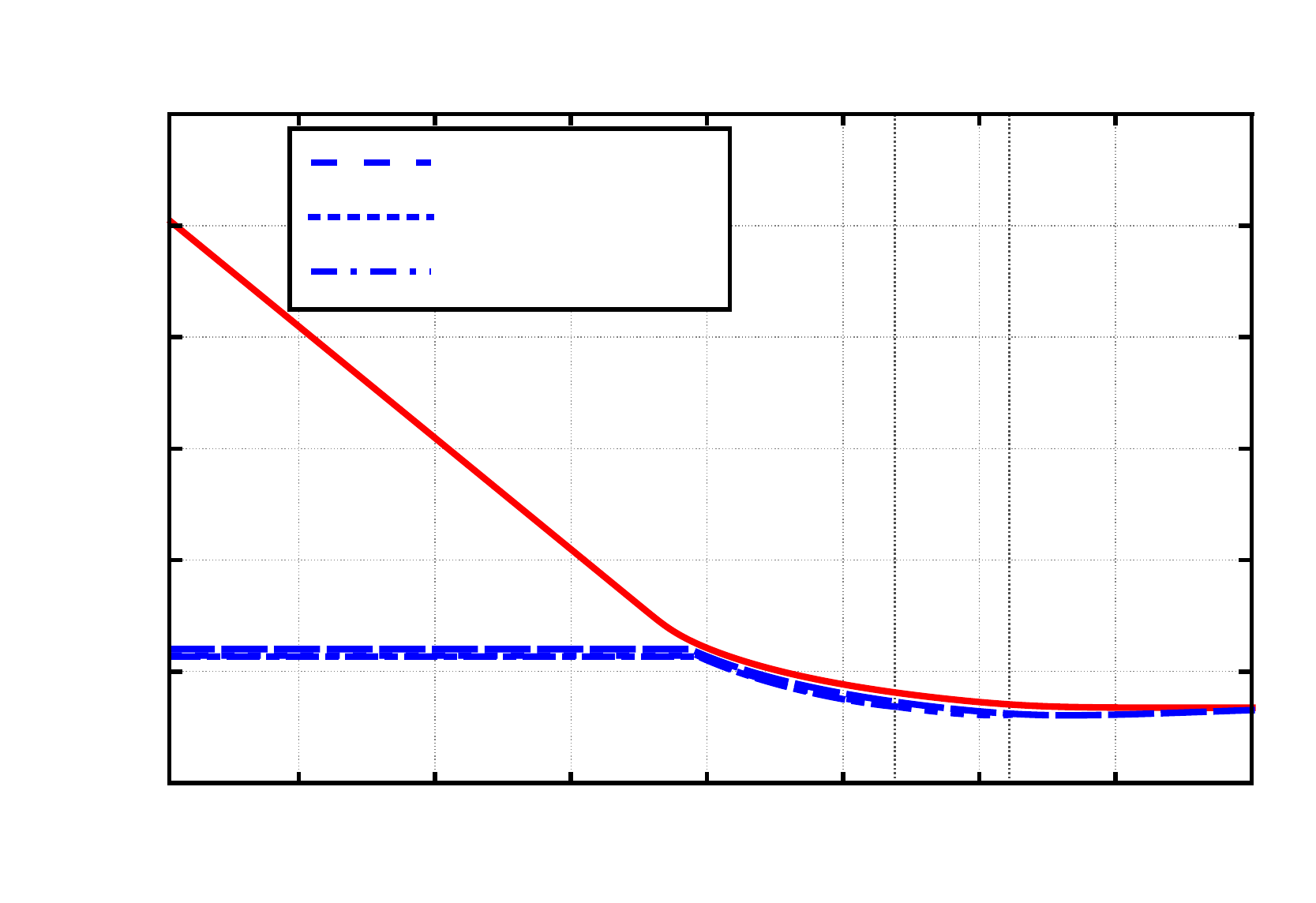}
        \put(35,100){\rotatebox{-38}{\footnotesize{Without Eve ($\eta_{zy}=0$)}}}
        \put(80,125.3){\small{$\eta_{zy}=0.5$}}
        \put(80,116){\small{$\eta_{zy}=0.9$}}
        \put(80,106.5){\small{$\eta_{zy}=0.99$}}
        \put(154.5,158){\small{$124.4$ dB}}
        \put(174.8,156){\vector(0,-1){19}}
        \put(126.5,148){\small{$107.6$ dB}}
        \put(154.9,147){\vector(0,-1){10}}
        \put(90,3){\small{Attenuation $\alpha$} [dB]}
        \put(26.5,12){\small{$0$}}
        \put(70.5,12){\small{$40$}}
        \put(117.5,12){\small{$80$}}
        \put(162,12){\small{$120$}}
        \put(209.5,12){\small{$160$}}
        \put(8,17.5){\small{$10^{-2}$}}
        \put(24,37){\small{$1$}}
        \put(14,56.3){\small{$10^2$}}
        \put(14,75.6){\small{$10^4$}}
        \put(14,94.9){\small{$10^6$}}
        \put(14,114.2){\small{$10^8$}}
        \put(10,133.5){\small{$10^{10}$}}
        \put(0,18){\rotatebox{90}{\footnotesize{Ave. photon no. $n^{\ast}_B$ [photons/pulse]}}}
      \end{overpic}
      \caption{Optimal average photon number $n^{\ast}_B$ of the received pulse for the secrecy rate $R_\mathrm{S}$ in Fig. \ref{zu35}.} \label{zu322}
    \end{figure}
    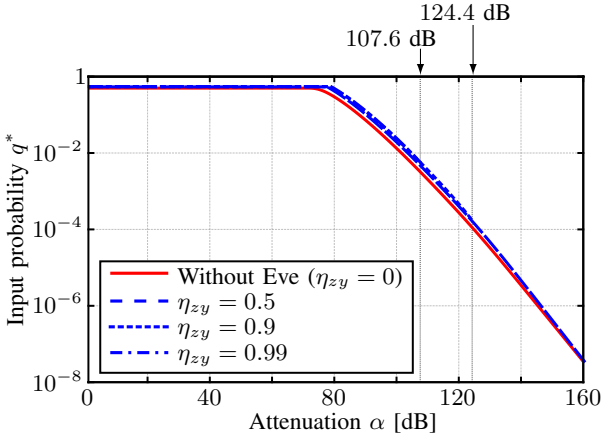
\begin{figure}[tp]
      \centering
      \begin{overpic}[width=8cm,bb=0 0 480 329]{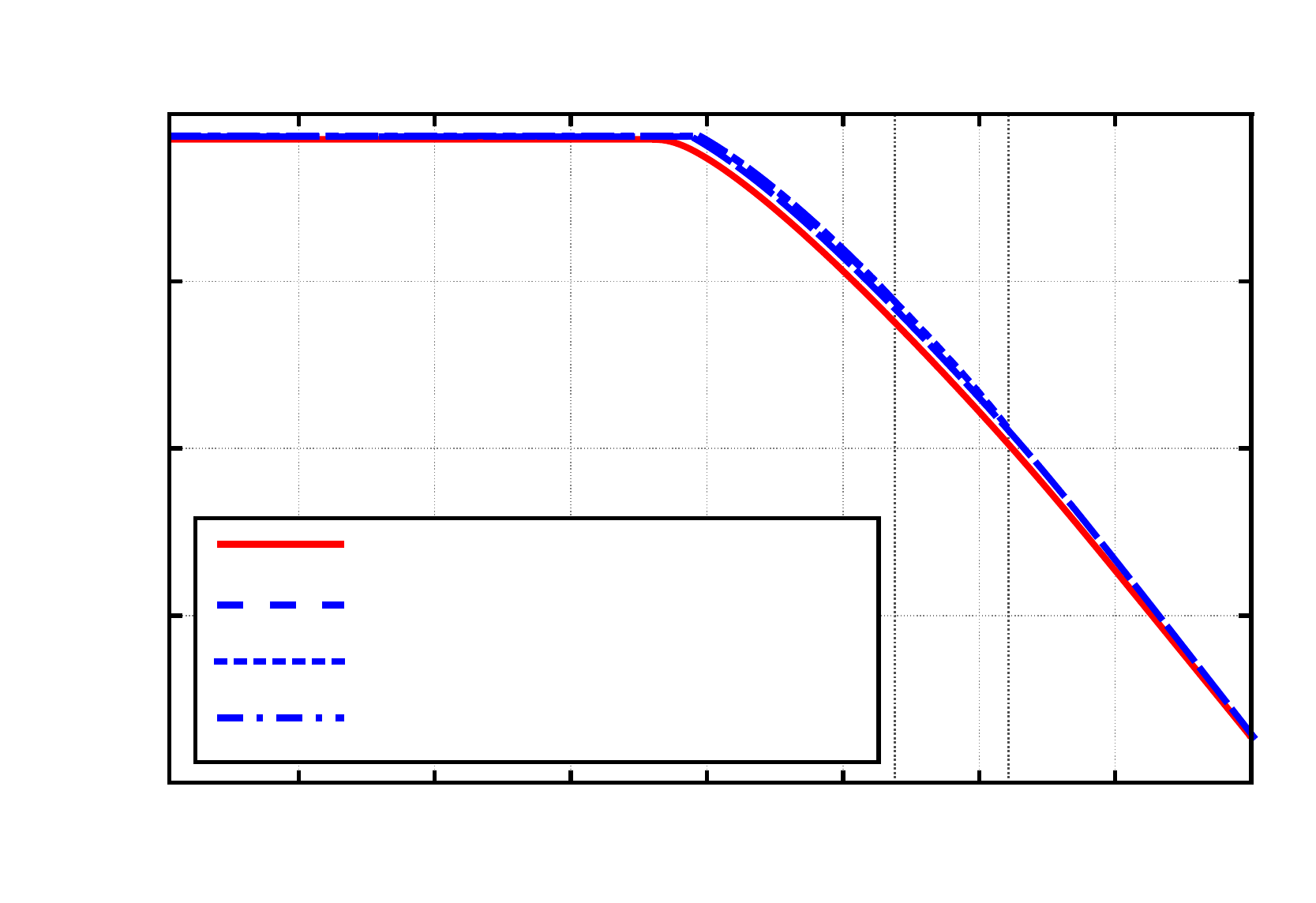}
        \put(63,58){\small{Without Eve ($\eta_{zy}=0$)}}
        \put(63,48.5){\small{$\eta_{zy}=0.5$}}
        \put(63,39){\small{$\eta_{zy}=0.9$}}
        \put(63,29){\small{$\eta_{zy}=0.99$}}
        \put(154.5,158){\small{$124.4$ dB}}
        \put(174.8,156){\vector(0,-1){19}}
        \put(126.5,148){\small{$107.6$ dB}}
        \put(154.9,147){\vector(0,-1){10}}
        \put(90,3){\small{Attenuation $\alpha$} [dB]}
        \put(26.5,12){\small{$0$}}
        \put(70.5,12){\small{$40$}}
        \put(117.5,12){\small{$80$}}
        \put(162,12){\small{$120$}}
        \put(209.5,12){\small{$160$}}
        \put(8,17.5){\small{$10^{-8}$}}
        \put(8,46.8){\small{$10^{-6}$}}
        \put(8,75.6){\small{$10^{-4}$}}
        \put(8,104.4){\small{$10^{-2}$}}
        \put(24,133.5){\small{$1$}}
        \put(-2,42){\small{\rotatebox{90}{Input probability $q^{\ast}$}}}
      \end{overpic}
      \caption{Optimal input probability $q^{\ast}$ for the secrecy rate $R_\mathrm{S}$ in Fig. \ref{zu35}.} \label{zu323}
    \end{figure} 
  
  In Fig. \ref{zu35}, we calculate the secrecy rate $R_\mathrm{S}$ as a function of attenuation $\alpha$ taking the above consideration into account.  
  As indicated in the figure, the secrecy rate decreases as the relative transmittance $\eta_{zy}$ gets close to $1$ which is the case where Eve receives the equal amount of power as Bob.  
  Compared to the channel capacity denoted by the solid line, we can observe some unique features of the secrecy rate $R_\mathrm{S}$ in terms of the dependence on attenuation $\alpha$.  
  First, in the noise-limited region, $R_\mathrm{S}$ decreases rapidly at a certain threshold point.  
  In this figure, $R_\mathrm{S}$ for $\eta_{zy}=0.9$ and $\eta_{zy}=0.99$ rapidly fall down to $0$ at around $\alpha=124.4$ dB and $107.6$ dB, respectively.  
  Second, if the available input power is fully used up in the loss-independent region, $R_\mathrm{S}$ rapidly falls down to $0$ as $\alpha$ decreases, 
  equivalently the distance between Alice and Bob gets shorter.  

  The optimal parameters $n^{\ast}_A$, $n^{\ast}_B$, and $q^{\ast}$ are depicted in Figs. \ref{zu32} - \ref{zu323}.  
  Interestingly enough, in contrast to the secrecy rate itself, the behaviors of these parameters seem to be irrespective to relative transmittance $\eta_{zy}$.  
  Fig. \ref{zu32} indicates that $n^{\ast}_A$ increases as $\alpha$ increases in both the loss-independent and noise-limited regions, 
  whereas $n^{\ast}_A$ for the channel capacity (solid line) stays constant.  
  This means that, for the secrecy rate, Alice should properly regulate the input power according to the distance between Alice and Bob.  
  Fig. \ref{zu322} shows the average photon number $n^{\ast}_B = \eta_y n^{\ast}_A$ of the received pulse.  
  As seen from the figure, in the loss-independent region, $n^\ast_B$ is kept unchanged even if the attenuation varies, 
  while this value slightly decreases but remains at few photons in the noise-limited region, 
  so that only Bob can discriminate the received signal from the noises but Eve should not so.  
  In contrast to the average photon number behaviors, the optimal input probability $q^\ast$ behaves in a way 
  showing no significant difference between the 
  secrecy rate \newline and channel capacity, as shown in Fig. \ref{zu323}.  

\subsection{Secrecy rate of wiretap channel coding and secure key rate of QKD}\label{sec34}
      \begin{figure}[tp]
        \centering
        \begin{overpic}[width=9cm,bb=0 0 507 296]{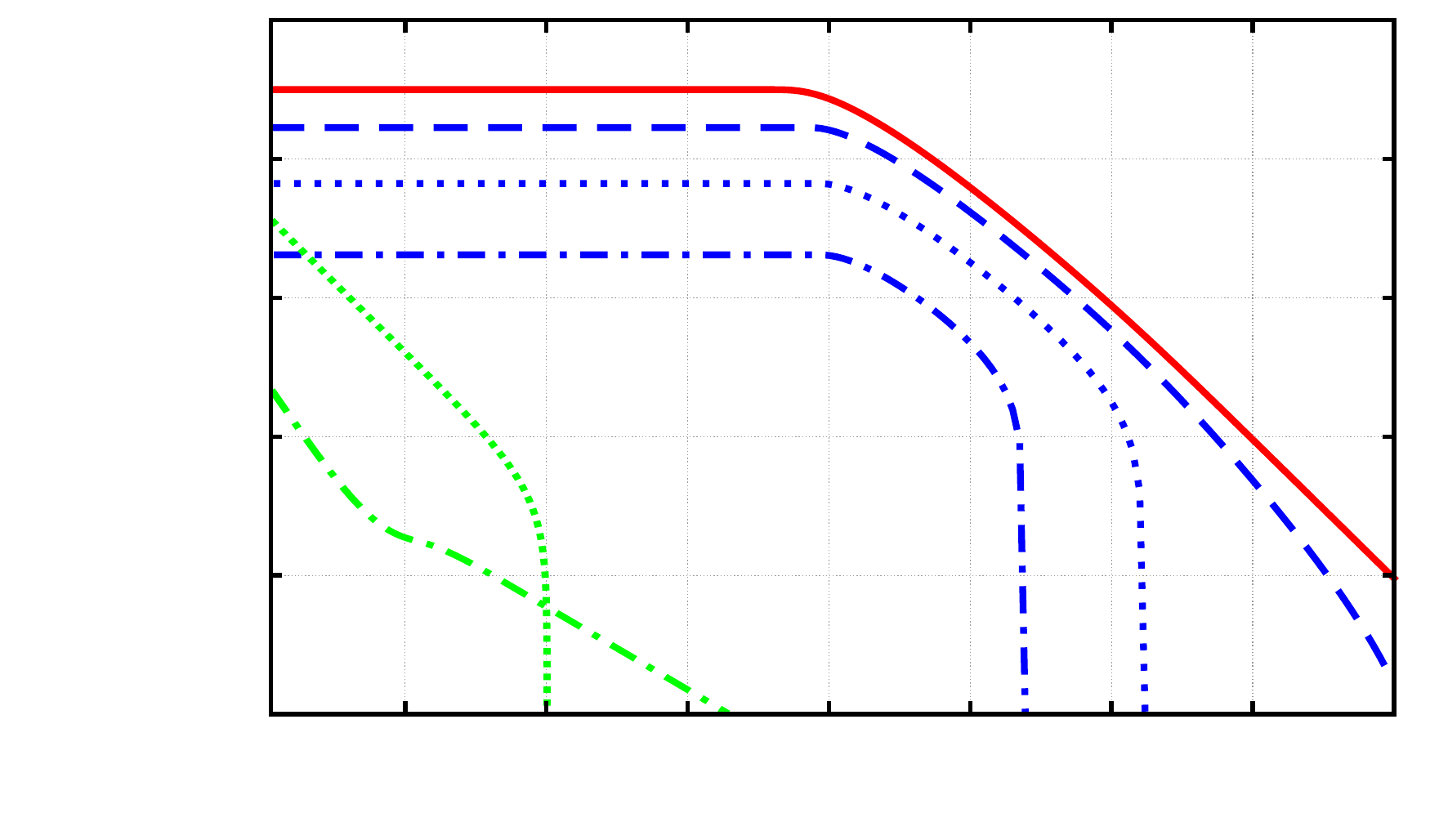}
          \put(110,3){\small{Attenuation $\alpha$ [dB]}}
          \put(46,14){\small{$0$}}
          \put(92.4,14){\small{$40$}}
          \put(143.3,14){\small{$80$}}
          \put(191.3,14){\small{$120$}}
          \put(241.3,14){\small{$160$}}
          \put(0,27){\rotatebox{90}{\small{Secure key rate}}}
          \put(11,45){\rotatebox{90}{\small{/Secrecy rate $R_\mathrm{S}$ [bps]}}}
          \put(43,21){\small{$1$}}
          \put(32,44){\small{$100$}}
          \put(32,68){\small{$10$k}}
          \put(34.5,93){\small{$1$M}}
          \put(25,118){\small{$100$M}}
          \put(30,143){\small{$10$G}}
          \put(68,136.5){\small{Channel capacity}}
          \put(63,99){\rotatebox{-47}{\small{Decoyed BB84}}}
          \put(60,65){\rotatebox{-30}{\small{Tele-amplified BB84}}}
          \put(111,120){\footnotesize{$\eta_{zy}\!=0.5$}}
          \put(110,121){\vector(-1,0.35){15}}
          \put(111,109){\footnotesize{$\eta_{zy}\!=0.9$}}
          \put(110,110){\vector(-1,0.37){15}}
          \put(111,90){\footnotesize{$\eta_{zy}\!=0.99$}}
          \put(110,91){\vector(-1,0.78){15}}
        \end{overpic}
        \caption{Secrecy rate $R_\mathrm{S}$ and secure key rate of QKD (BB84 \cite{BB} protocol).  
        Parameters for wiretap channel: $P=10$ mW, $\lambda_y=10$ kcps, $\lambda_z=1$ cps, $\Delta=1$ ns.  
        Parameters for QKD: pulse generation rate = $1$ GHz, DCR of a detector = $100$ cps.  }
        \label{zu31}
      \end{figure}
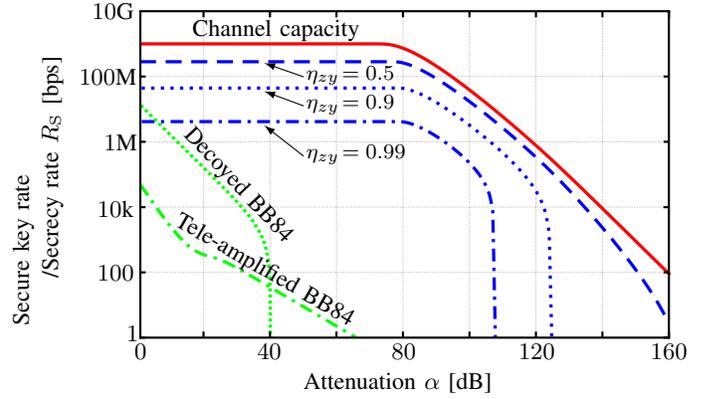
    
    In Fig. \ref{zu31}, we show simultaneously in the same graph the secrecy rate obtained in the previous subsection and the secure key rate of QKD schemes.  
    The secrecy rate measures a message rate of wiretap channel coding for one-way transmission, 
    while the secure key rate of QKD does a rate of key exchange with quantum channel and an authenticated public (classical) channel.
    The two schemes are based on different assumptions on Eve. Fig. \ref{zu31} aims at showing how we can increase the rate and distance of FSO links with ITS 
    by compromising the assumption on Eve within reasonable practical conditions.  
    
    The curve labeled with ``decoyed BB84" shows a theoretical prediction of the secure key rate via BB84 \cite{BB} employing the decoy-pulse method \cite{decoy}.  
    Here, we assume an ideal linear attenuation channel and a single photon detector with a repetition rate of $1$ GHz and a DCR of $100$ cps, which is a typical DCR for the current QKD systems.  
    This figure indicates that the secure key rate rapidly falls down at a distance of $40$ dB attenuation, which is roughly the best link budget for a low earth orbit (LEO) to ground distance in optical space communications \cite{Sasakiinvite}.  
    Although quantum relay based on tele-amplification \cite{teleamp} has been proposed for extending a QKD distance (see the curve labeled with ``tele-amplified BB84"), 
    the secure key rate is always sacrificed while extending the transmission distance.  
    
    On the other hand, as shown in Fig. \ref{zu31}, the secrecy rate $R_\mathrm{S}$ (blue lines) can cover a wider range in which QKD hardly generates the secure key even for 
    the relative transmittance as high as $\eta_{zy}=0.99$ 
    for the case where Eve can obtain $99\%$ as much power as Bob. 
    Fig. \ref{zu31} shows FSO links with ITS is possible even at $\alpha=80$ dB which roughly corresponds to the best link budget for a GEO-ground distance.  
    Consequently, wiretap channel coding is potentially a promising candidate for realization of the global scale secure network based on FSO communications.  
    
\section{Secrecy capacity} \label{sec4}
  In this section, we extend the analysis to full optimization of secrecy rate by introducing the auxiliary random variable $V$ at Alice, 
  as was formulated by Csisz\'ar and K\"orner \cite{csiskor} and study the secrecy capacity.  
  This scheme requires us to concatenate an additional channel $P_{X|V}$ to the main channel $W_B$ and the wiretapper channel $W_E$, respectively.  
  We reformulate the previous tools, present numerical results, and clarify the functional meaning and quantitative effects of the auxiliary random variable $V$. 
  
\subsection{Power constraint and channel matrices}
    \begin{figure}[tb]
      \centering
      \begin{picture}(200,270)(0,0)
        \thicklines
        \multiput(87,237.4)(2,-3.50){27}{\color{Salmon}{\line(1,-1.75){1}}}
        \multiput(87,206.8)(2,-1.21){27}{\color{Salmon}{\line(1,-0.61){1}}}
        \put(87,237.4){\color{Red}{\line(1,-1.18){53.7}}}
        \put(87,206.8){\color{Red}{\line(1,-1.18){53.7}}}
        \multiput(87,237.4)(2,-1.14){27}{\color{Gray}{\line(1,-0.57){1}}}
        \multiput(87,206.8)(2,1.14){27}{\color{Gray}{\line(1,0.57){1}}}
        \put(87,237.4){\line(1,0){53.7}}
        \put(87,206.8){\line(1,0){53.7}}
        \multiput(28.5,206.8)(2,1.16){26}{\color{Gray}{\line(1,0.58){1}}}
        \multiput(28.5,237.4)(2,-1.16){26}{\color{Gray}{\line(1,-0.58){1}}}
        \put(28.5,237.4){\line(1,0){52.5}}
        \put(28.5,206.8){\line(1,0){52.5}}
        \put(-5,235.1){\small{$v=0$}}
        \put(25.5,237.4){\circle{6}}
        \put(-5,204.7){\small{$v=1$}}
        \put(25.5,206.8){\circle{6}}
        \put(70,215){\small{$a$}}
        \put(52,208){\small{$b$}}
        \put(45,243){\small{$P_{X|V}$}}
        \put(69,245){\small{$x=0$}}
        \put(84,237.4){\circle{6}}
        \put(69,192){\small{$x=1$}}
        \put(84,206.8){\circle{6}}
        \put(108,243){\small{$W_B$}}
        \put(108,160){\small{\textcolor{Red}{$W_E$}}}
        \put(150,235.1){\small{$y=0$}}
        \put(143.7,237.4){\circle{6}}
        \put(150,204.7){\small{$y=1$}}
        \put(143.7,206.8){\circle{6}}
        \put(150,171.8){\small{\textcolor{Red}{$z=0$}}}
        \put(143.7,174){\color{red}{\circle{6}}}
        \put(150,141.1){\small{\textcolor{Red}{$z=1$}}}
        \put(143.7,143.3){\color{red}{\circle{6}}}
        \put(78,260){$(a)$}

        \multiput(28.5,97.4)(2,-1.672){56}{\color{Salmon}{\line(1,-0.836){1}}}
        \multiput(28.5,66.8)(2,-0.584){56}{\color{Salmon}{\line(1,-0.292){1}}}
        \put(28.5,97.4){\color{red}{\line(1,-0.564){112.5}}}
        \put(28.5,66.8){\color{red}{\line(1,-0.564){112.5}}}
        \multiput(28.5,97.4)(2,-0.54){56}{\color{Gray}{\line(1,-0.27){1}}}
        \multiput(28.5,66.8)(2,0.54){56}{\color{Gray}{\line(1,0.27){1}}}
        \put(28.5,97.4){\line(1,0){112.5}}
        \put(28.5,66.8){\line(1,0){112.5}}
        \put(-5,95.1){\small{$v=0$}}
        \put(25.5,97.4){\circle{6}}
        \put(-5,64.7){\small{$v=1$}}
        \put(25.5,66.8){\circle{6}}
        \put(78,103){\small{$W^+_B$}}
        \put(78,22.5){\small{\textcolor{red}{$W^+_E$}}}
        \put(150,95.1){\small{$y=0$}}
        \put(143.5,97.4){\circle{6}}
        \put(150,64.7){\small{$y=1$}}
        \put(143.5,66.8){\circle{6}}
        \put(118,76){\small{$a^+_y$}}
        \put(118,58){\small{$b^+_y$}}
        \put(150,31.8){\small{\textcolor{red}{$z=0$}}}
        \put(143.5,34){\color{red}{\circle{6}}}
        \put(150,1.1){\small{\textcolor{red}{$z=1$}}}
        \put(143.5,3.3){\color{red}{\circle{6}}}
        \put(78,120){$(b)$}
        \put(118,26.5){\small{\textcolor{red}{$a^+_z$}}}
        \put(118,4){\small{\textcolor{red}{$b^+_z$}}}
      \end{picture}
      \caption{(a) Channel diagram of the wiretap channel with an auxiliary random channel $P_{X|V}$. 
      (b) Channel diagram of the concatenated channels $W^+_B, W^+_E$.  } \label{transitionprobabilityconc}
    \end{figure}
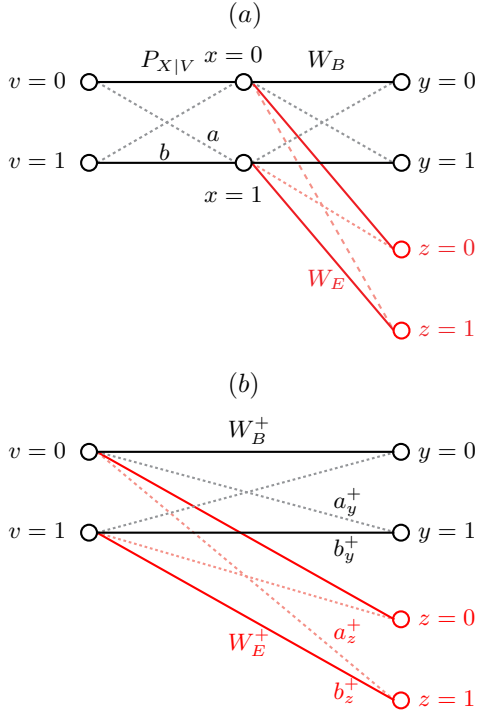
  Similarly to the model in Section \ref{sec2}, Alice generates the on- and off-signals corresponding to encoding symbols ``1" and ``0" with probabilities $q$ and $1-q$, respectively.  
  Then, Alice inputs the sequences into the wiretap channel with picking a symbol and flipping it randomly.  
  Here, the encoding and input symbols are modeled by the auxiliary random variable $V$ and the input random variable $X$, respectively  
  (more formally, the random variables form a Markov chain $V \haihun X \haihun YZ$).  
  Since the number taken by the elements of the auxiliary random variable $V$ need not exceed that of the input random variable $X$ \cite{CKbook}, 
  we consider the case where $V$, $X$, $Y$, and $Z$ are all binary, as illustrated in Fig. \ref{transitionprobabilityconc}(a).  
  The auxiliary channel $P_{X|V}$ from $V$ to $X$ can be modeled by the channel matrix elements with any constants $0<a,b<1$ as follows:
    \begin{align}
      P_{X|V}(1|0) = a, \quad P_{X|V}(1|1) = b. \label{pxv11}
    \end{align}
  
  Then, the probability of the input pulse into the wiretap channel is $q^{+} \equiv (1-q)a +q (1-b)$ in each time slot.  
  Thus, the power constraint of (\ref{powerconst}) which is imposed on $X$ is rewritten as
    \begin{equation}
      q^{+} \frac{n_A h f_0}{\Delta} \leq P. \label{powerconstV}
    \end{equation}
  
  Effectively, we can consider the concatenated channels $W^+_B$ from $V$ to $Y$ and $W^+_E$ from $V$ to $Z$ as shown in Fig. \ref{transitionprobabilityconc}(b).  
  Given the main channel $W_B$ and the auxiliary channel $P_{X|V}$, the conditional probability of the concatenated channel $W^+_B$ can be written as  
    \begin{equation}
      W^+_B(y|v) = \sum_{x \in \{0,1\}} W_B(y|x) P_{X|V}(x|v).  
    \end{equation}
  The channel matrix of $W^+_B$ is given by 
    \begin{align}
      W^+_B(1|0) = (1-a)a_y + a b_y \equiv a^{+}_y,& \quad W^+_B(0|0) = 1-a^{+}_y, \non \\
      W^+_B(1|1) = (1-b)a_y + b b_y \equiv b^{+}_y,& \quad W^+_B(0|1) = 1-b^{+}_y. \non
    \end{align}
  Likewise, the channel matrix of the concatenated channel $W^+_E$ is given by
    \begin{align}
      W^+_E(1|0) = (1-a)a_z + a b_z \equiv a^{+}_z,& \quad W^+_E(0|0) = 1-a^{+}_z, \non \\
      W^+_E(1|1) = (1-b)a_z + b b_z \equiv b^{+}_z,& \quad W^+_E(0|1) = 1-b^{+}_z. \non
    \end{align}

\subsection{Secrecy capacity}
    \begin{figure}[tp]
      \centering
      \begin{overpic}[width=8cm,bb = 0 0 475 291]{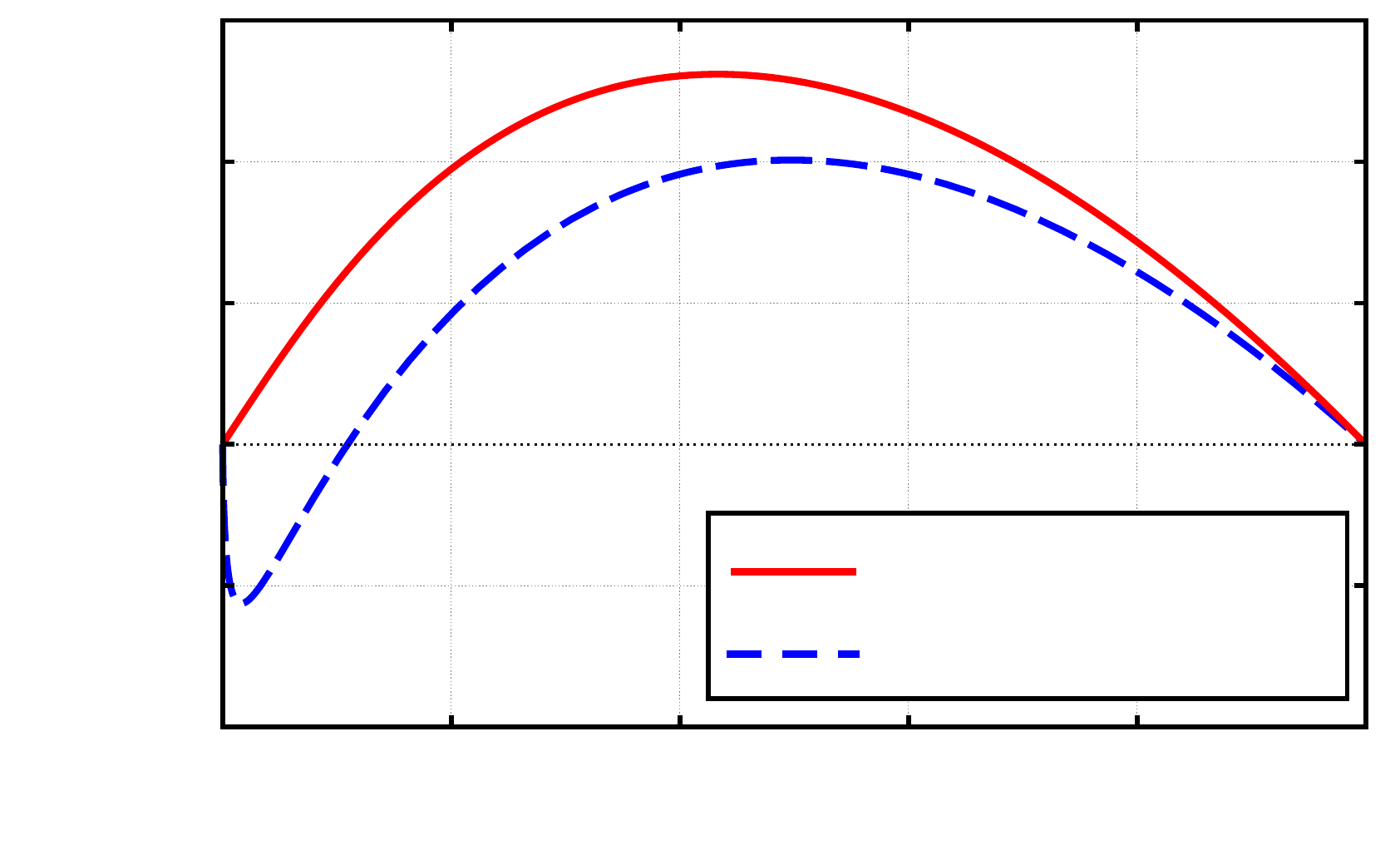}
        \put(93,2){\small{Input probability $q$}}
        \put(34,12){\small{$0$}}
        \put(69.5,12){\small{$0.2$}}
        \put(107.5,12){\small{$0.4$}}
        \put(145,12){\small{$0.6$}}
        \put(183,12){\small{$0.8$}}
        \put(221,12){\small{$1.0$}}
        \put(10,30){\rotatebox{90}{\small{Value of function [kbps]}}}
        \put(18,19.5){\small{$-40$}}
        \put(18,40){\small{$-20$}}
        \put(30,63.5){\small{$0$}}
        \put(25,87){\small{$20$}}
        \put(25,111){\small{$40$}}
        \put(25,133){\small{$60$}}
        \put(145,45){\footnotesize{$\displaystyle \max_{a,b}f^+_{BE}(q,n_A,a,b)$}}
        \put(145,29){\footnotesize{$\displaystyle f_{BE}(q,n_A)$}}
      \end{overpic}
      \caption{
      Comparison of $\max_{a,b}f^+_{BE}(q,n_A,a,b)$ with $f_{BE}(q,n_A)$ varying the input probability $q$.  
      Parameters: $n_B = 3.2 \times 10^{-3}$ photons/pulse, $\eta_{zy} = 0.95$, $\lambda_y=10$ kcps, $\lambda_z=1$ cps, $\Delta=1$ ns.  
      } \label{diffmutualconcatonation}
    \end{figure}
    \begin{figure*}[tp]
      \centering
      \begin{overpic}[width=16cm,bb =  0 0 955 327]{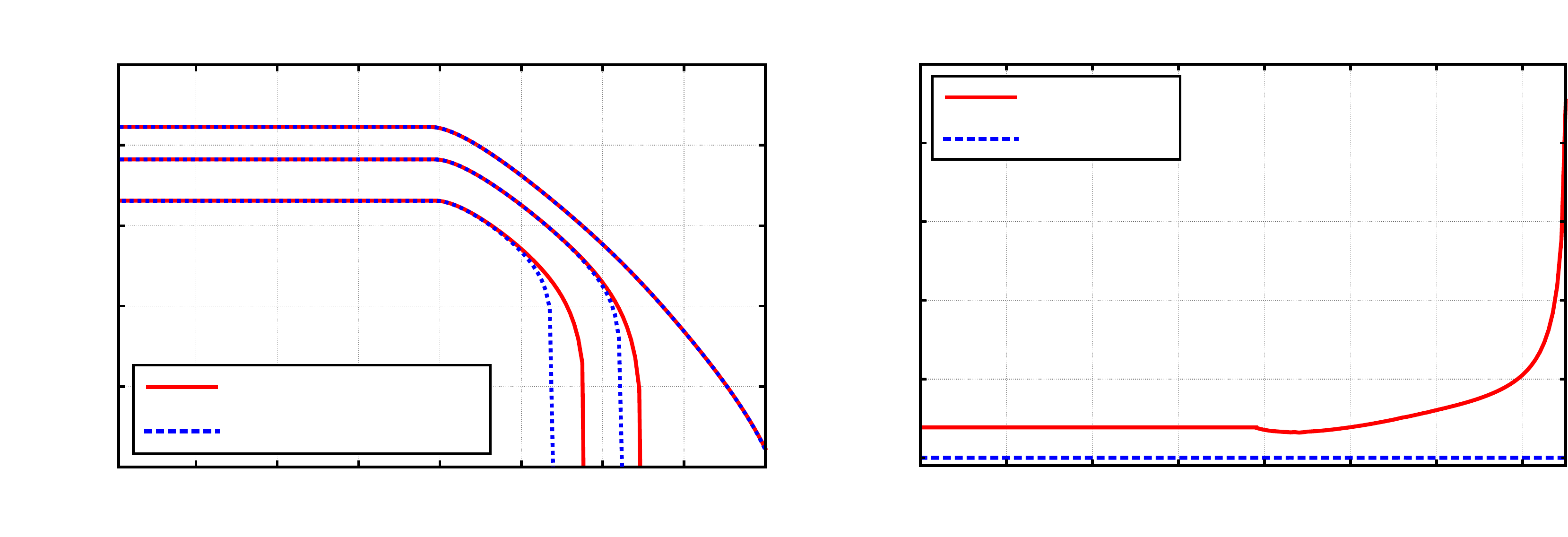}
        \put(94,2){\small{Attenuation $\alpha$ [dB]}}
        \put(32,12){\small{$0$}}
        \put(76,12){\small{$40$}}
        \put(123,12){\small{$80$}}
        \put(168,12){\small{$120$}}
        \put(215,12){\small{$160$}}
        \put(-1,37){\rotatebox{90}{\small{Secrecy capacity [bps]}}}
        \put(27,18){\small{$1$}}
        \put(17,41){\small{$100$}}
        \put(17,64){\small{$10$k}}
        \put(19,88){\small{$1$M}}
        \put(10,110){\small{$100$M}}
        \put(15.5,134){\small{$10$G}}
        \put(66,40){\small{$C_\mathrm{S}$ (with $P_{X|V}$)}}
        \put(66,28){\small{$R_\mathrm{S}$ (without $P_{X|V}$)}}
        \put(55,125){\small{$\eta_{zy}=0.5$}}
        \put(53,126){\vector(-1,-0.6){10}}
        \put(55,102){\small{$\eta_{zy}=0.9$}}
        \put(53,102.5){\vector(-1,0.6){11}}
        \put(55,86){\small{$\eta_{zy}=0.99$}}
        \put(53,86){\vector(-1,0.9){12}}
        \put(122.5,145){\small{(a)}}
        \put(333,2){\small{Attenuation $\alpha$ [dB]}}
        \put(262,12){\small{$40$}}
        \put(313,12){\small{$60$}}
        \put(362.5,12){\small{$80$}}
        \put(410.2,12){\small{$100$}}
        \put(248,36){\rotatebox{90}{\small{Cross-over probabilities}}}
        \put(260,20){\small{$0$}}
        \put(260,43){\small{$1$}}
        \put(260,66){\small{$2$}}
        \put(260,88){\small{$3$}}
        \put(260,112){\small{$4$}}
        \put(260,135){\small{$5$}}
        \put(268,139){\small{$\times 10^{-3}$}}
        \put(300,126){\small{$P_{X|V}(1|0)$}}
        \put(300,114){\small{$P_{X|V}(0|1)$}}
        \put(161,30){\color{red}{\vector(1,0){8}}}
        \put(358,145){\small{(b)}}
      \end{overpic}
      \caption{
      (a) Secrecy capacity $C_\mathrm{S}$ as a function of attenuation $\alpha = - \log_{10} \eta_y$.    
      Also shown for comparison are the secrecy rate $R_\mathrm{S}$.  
      Parameters: $P=10$ mW, $\lambda_y=10$ kcps, $\lambda_z=1$ cps, $\Delta=1$ ns. 
      (b) Optimal cross-over probabilities $P_{X|V}(1|0)=a$ and $P_{X|V}(0|1)=1-b$ for $\eta_{zy} = 0.99$.}
      \label{zu37}
    \end{figure*}
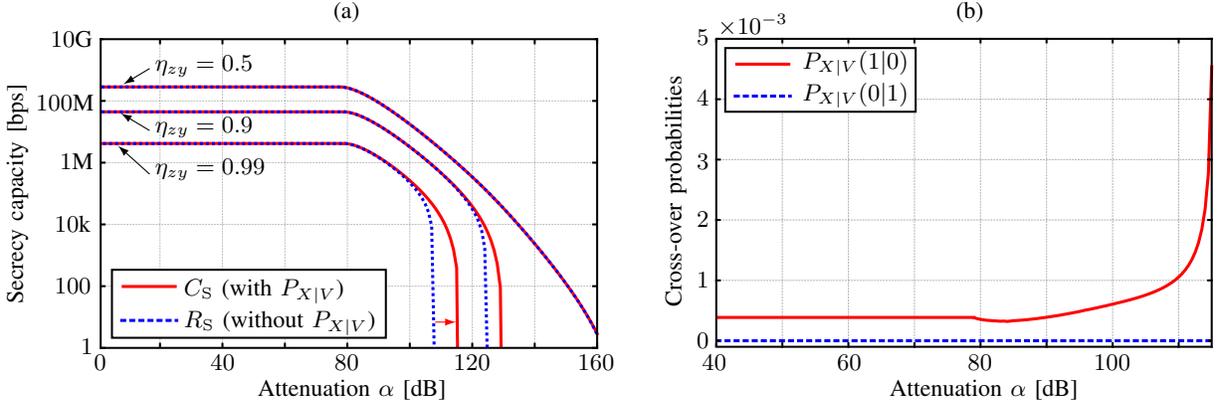
  With the channel matrices given in the previous subsection, the secrecy capacity $C_{\mathrm{S}}$ is defined and computed as the simultaneous
  optimization over $q$, $n_A$, $a$, and $b$:
   \begin{equation}
      C_{\mathrm{S}} = \max_{q,n_A,a,b} f^+_{BE}(q,n_A,a,b), \label{secrecyV} 
    \end{equation}
  where the function $f^+_{BE}(q,n_A,a,b)$ is defined to be
    \begin{align}
      f^+_{BE}(q,n_A,a,b) \equiv f^+_B(q,n_A,a,b) - f^+_E(q,n_A,a,b),
    \end{align}
  and $f^+_B(q,n_A,a,b)$ and $f^+_E(q,n_A,a,b)$ are the mutual informations:
    \begin{align}
      f^+_B(q,n_A,a,b) \equiv & h_2 ((1-q)a^+_y + q(1- b^+_y)) \non \\
                              & \quad - (1-q) h_2 (a^+_y) - q h_2 (b^+_y), \\
      f^+_E(q,n_A,a,b) \equiv & h_2((1-q)a^+_z + q(1- b^+_z)) \non \\
                              & \quad - (1-q)h_2(a^+_z) - qh_2(b^+_z),
    \end{align}
  where $n_A$ intervenes through $a_y, b_y, a_z, b_z$.  
  
  In Fig. \ref{diffmutualconcatonation}, we compare the function $\max_{a,b}f^+_{BE}(q,n_A,a,b)$ (solid line) optimized over $a$ and $b$ with 
  $f_{BE}(q,n_A)$ (dashed line) for a wiretap channel which is not more capable.  
  The figure indicates that $\max_{a,b}f^+_{BE}(q,n_A,a,b)$ is strictly positive for any input probability $q \in \{0,1\}$ whereas $f_{BE}(q,n_A)$ becomes negative for small $q$.  
  Moreover, $\max_{a,b}f^+_{BE}(q,n_A,a,b)$ is larger than $f_{BE}(q,n_A)$ at any input probability $q \in \{0,1\}$.  
  The extension in the transmission distance which will be shown later should be attributed to this increase of the value caused by the auxiliary random variable $V$.  
    
\subsection{Numerical evaluation}
  In this subsection, we numerically demonstrate the improvement of transmission distance due to the concatenation of the auxiliary channel $P_{X|V}$.  
  In Fig. \ref{zu37}(a), we compare the secrecy capacity $C_{\mathrm{S}}$ (solid lines)  based on (\ref{secrecyV}) 
  with the secrecy rate $R_{\mathrm{S}}$ (dashed lines) based on (\ref{secrecyX}) which was investigated in Sections \ref{sec2} and \ref{sec3} (see Fig. \ref{zu35}).  
  According to the figure, the auxiliary random variable $V$ brings about the improvement of transmission distance in the noise limited region, e.g., 
  for $\eta_{zy}=0.99$, the attenuation $\alpha$ at which the secrecy rate sharply falls is improved by $6$ dB, which is equivalent to $40$\% extension of the transmission distance.  
  This effect becomes significant for larger values of the relative transmittance $\eta_{zy}$.  
  
  Fig. \ref{zu37}(b) shows the optimal cross-over probabilities $P_{X|V}(1|0)=a$ and $P_{X|V}(0|1)=1-b$ for $\eta_{zy} = 0.99$ in Fig. \ref{zu37}(a).  
  Here, $P_{X|V}(1|0)$ is the probability of flipping ``0" (off-signal) into ``1" (on-signal) and $P_{X|V}(0|1)$ is vice versa.  
  As seen in Fig. \ref{zu37}(b), $P_{X|V}(1|0)$ is non-zero and increases drastically in the noise-limited region, whereas $P_{X|V}(0|1)$ stays $0$.
  
  The effect of the auxiliary randomness generated at the sender on the performance has been investigated especially in the multiple receivers scenario \cite{artificial, SINR, visibleindoor}, 
  namely, the artificial noise is created such that it degrades Eve's channel but does not affect the main channel through the use of the interference effect among the receivers.  
  In contrast to such studies, Fig. \ref{zu37}(b) reveals that the addition of the random pulses has a crucial role in the proposed method.  
  In our case, Eve who may have the less noisy detector than Bob can be further deceived by the dummy pulses which act as extra noises, and the performance is enhanced.  
  This means that the proposed method bears a remarkable resemblance to the decoy method employed in BB84 \cite{decoy}.  
  In this method, Alice varies the average photon number of each signal pulse randomly among the prescribed levels, thus Eve is prevented from wiretapping the signal pulses,
  and the security and the transmission distance is boosted.  

\section{Finite length analysis} \label{sec5}
\subsection{Formulation}
    \begin{figure}[t!]
      \centering
      \begin{picture}(225,70)(0,0)
        \put(22,8){\small{\textcolor{blue}{Message bit $m$}}}
        \linethickness{1pt}
        \multiput(0,31)(15,0){7}{\color{blue}{\framebox(14,14){}}}
        \multiput(105,31)(15,0){4}{\color{red}{\framebox(14,14){}}}
        \multiput(165,31)(15,0){4}{\framebox(14,14){}}
        \put(-0.5,28){\line(0,-1){20}}
        \put(104.5,28){\line(0,-1){20}}
        \put(164.5,28){\line(0,-1){20}}
        \put(224.5,28){\line(0,-1){20}}
        \put(-0.5,48){\line(0,1){20}}
        \put(224.5,48){\line(0,1){20}}
        \put(3,18){\color{blue}{\vector(1,0){101}}}
        \put(101,18){\color{blue}{\vector(-1,0){101}}}
        \put(108,8){\small{\textcolor{red}{Randomness}}}
        \put(127,-2){\small{\textcolor{red}{bit $l$}}}
        \put(108,18){\color{red}{\vector(1,0){56}}}
        \put(161,18){\color{red}{\vector(-1,0){56}}}
        \put(168,18){\vector(1,0){56}}
        \put(221,18){\vector(-1,0){56}}
        \put(172,8){\small{Redundant}}
        \put(190,-2){\small{bit}}
        \put(3,58){\vector(1,0){221}}
        \put(221,58){\vector(-1,0){221}}
        \put(90,62){\small{Code length $n$}}
      \end{picture}
      \caption{Conceptual codeword structure of a wiretap channel code.} \label{zu11}
    \end{figure}
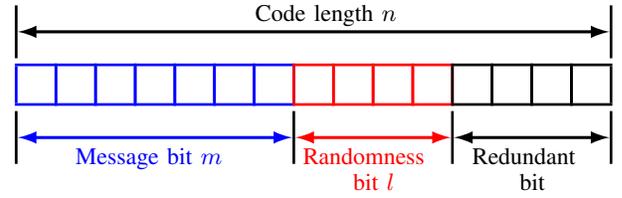
    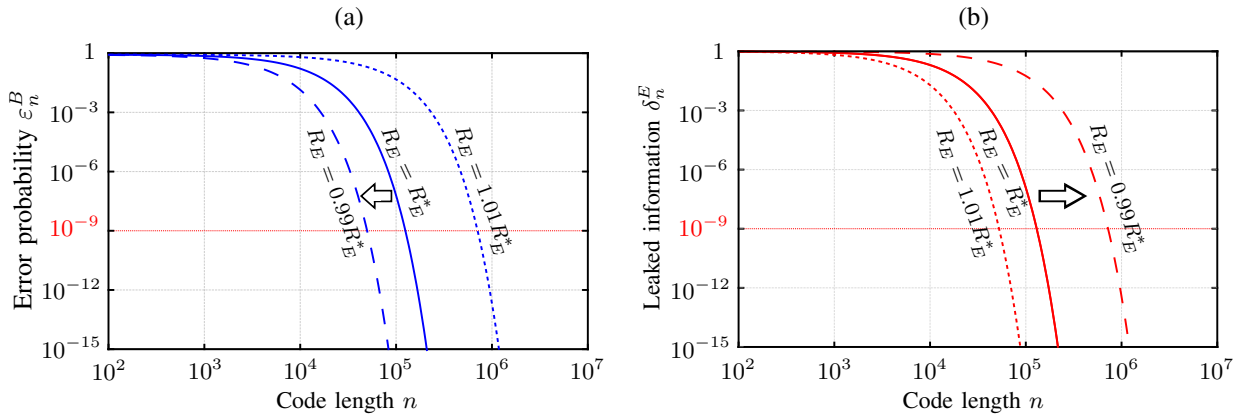
\begin{figure*}[tp]
      \centering
      \begin{overpic}[width=16cm,bb =0 0 990 335]{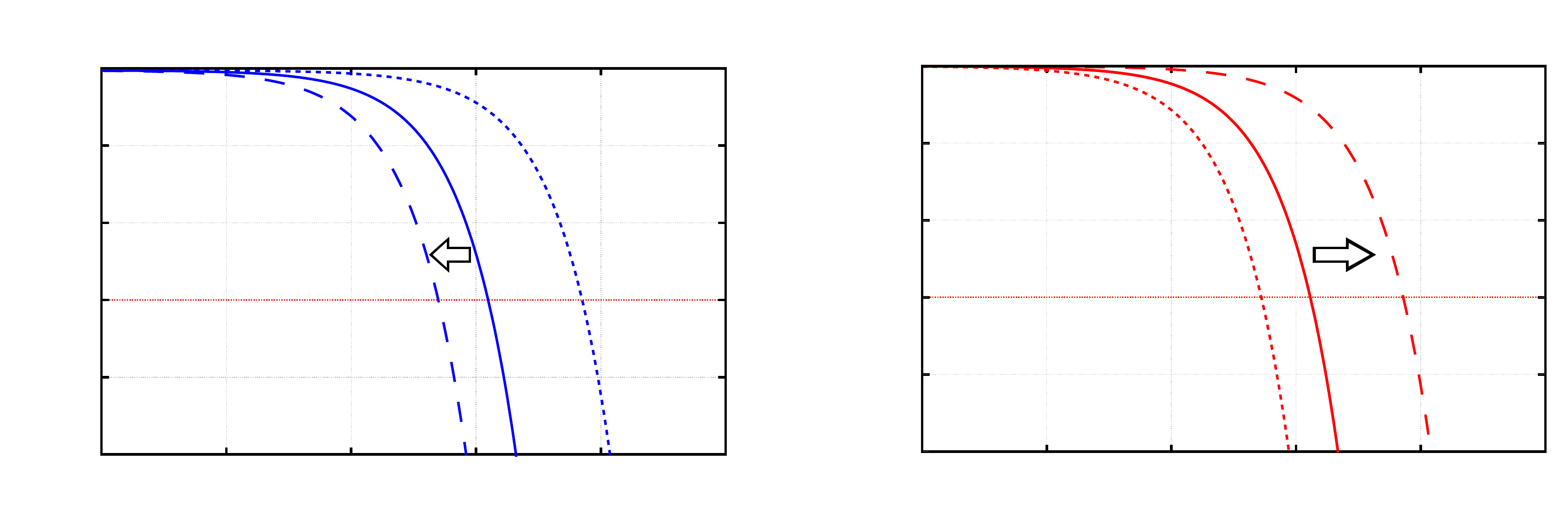}
        \put(23,10.5){\small{$10^2$}}
        \put(59.5,11){\small{$10^3$}}
        \put(95.5,11){\small{$10^4$}}
        \put(131.5,11){\small{$10^5$}}
        \put(168,11){\small{$10^6$}}
        \put(204,11){\small{$10^7$}}
        \put(93,0){\small{Code length $n$}}
        \put(4,19){\small{$10^{-15}$}}
        \put(4,41){\small{$10^{-12}$}}
        \put(7,65){\small{\textcolor{red}{$10^{-9}$}}}
        \put(7,86){\small{$10^{-6}$}}
        \put(7,108){\small{$10^{-3}$}}
        \put(20,132){\small{$1$}}
        \put(-8,37){\rotatebox{90}{Error probability $\vep^B_n$}}
        \put(130,105){\rotatebox{-65}{\footnotesize{$R_E = R^{\ast}_E$}}}
        \put(102,105){\rotatebox{-70}{\footnotesize{$R_E = 0.99R^{\ast}_E$}}}
        \put(158,106){\rotatebox{-70}{\footnotesize{$R_E = 1.01R^{\ast}_E$}}}
        \put(115,145){(a)}
        \put(261,11){\small{$10^2$}}
        \put(297.5,11){\small{$10^3$}}
        \put(333.5,11){\small{$10^4$}}
        \put(369.5,11){\small{$10^5$}}
        \put(406,11){\small{$10^6$}}
        \put(442,11){\small{$10^7$}}
        \put(330,0){\small{Code length $n$}}
        \put(242,19){\small{$10^{-15}$}}
        \put(242,41){\small{$10^{-12}$}}
        \put(245,65){\small{\textcolor{red}{$10^{-9}$}}}
        \put(245,86){\small{$10^{-6}$}}
        \put(245,108){\small{$10^{-3}$}}
        \put(258,132){\small{$1$}}
        \put(230,37){\rotatebox{90}{\small{Leaked information $\delta^E_n$}}}
        \put(356,105){\rotatebox{-65}{\footnotesize{$R_E = R^{\ast}_E$}}}
        \put(398,107){\rotatebox{-71}{\footnotesize{$R_E = 0.99R^{\ast}_E$}}}
        \put(340,105){\rotatebox{-70}{\footnotesize{$R_E = 1.01R^{\ast}_E$}}}
        \put(351,145){(b)}
      \end{overpic}
      \caption{Code length dependence of (a) error probability $\vep^B_n$ and (b) leaked information $\delta^E_n$.  
      The coding rate is fixed as to be $R^{\ast}_B = 0.5 R_{\mathrm{S}} = 22.1$ Mbps (see Fig. \ref{zu35}).  
      The arrows denote the change of the code length dependence when the randomness rate decreases by $1$\% from $R^{\ast}_E = 0.641$ Gbps.  } \label{zu52}
    \end{figure*}
    \begin{table*}[tp]
      \renewcommand{\arraystretch}{1.3}
      \caption{Rates and exponents for Fig. \ref{zu52}}
      \label{table1}
      \centering
      \begin{tabular}{c|cccc}
      \hline
      & $R_E$ [Gbps]& $R^{\ast}_B+R_E$ [Gbps]& $F_c(q, R_B, R_E)$ & $H_c(q,R_E)$ \\
      \hline
      $R_E = R^{\ast}_E$ & $0.641 $ & $0.663$ & $1.59 \times 10^{-4}$ & $1.59 \times 10^{-4}$ \\
      $R_E = 0.99 R^{\ast}_E$ & $0.634$ & $0.656$ & $4.00 \times 10^{-4}$ & $0.29 \times 10^{-4}$ \\
      $R_E = 1.01 R^{\ast}_E$ & $0.647$ & $0.669$ & $0.28 \times 10^{-4}$ & $3.94 \times 10^{-4}$ \\
      \hline
      \end{tabular}
    \end{table*}
  Although the secrecy capacity investigated in the last section is considered as a reasonable benchmark of the system, 
  it concerns only the achievable rate in the asymptotic limit at code length $n \to \infty$ and contains no information about practical code construction of finite length $n$.  
  In this section, in order to estimate required resources for the given levels of reliability for Bob and secrecy against Eve, 
  we introduce a stronger characterization showing how fast the decoding error probability and the leaked information decrease.  

  As depicted schematically in Fig. \ref{zu11}, a wiretap channel code consists of three types of bits, i.e., 
  $m$ bits conveying the confidential information, $l$ bits as the random dummy information to deceive Eve, and $n-m-l$ redundant bits to perform error correction.  
  Here, let $R_B = m/n$ and $R_E = l/n$ be the coding rate and the randomness rate, respectively.  

  For such a code, we introduce the measures on the decoding error probability $\vep^B_n$ and the leaked information $\delta^E_n$.  
  Given an output via the main channel due to message $i$ from the message set $\cM_n$, Bob computes an estimate $\hat{i}$ of message $i$ with his decoder.  
  The decoding error probability $\vep^B_n$ is measured as
    \begin{equation}
      \vep^B_n \equiv \frac{1}{|\cM_n|} \sum_{i \in \cM_n} \Pr\{i \neq \hat{i}\}, \label{errorprob}
    \end{equation}
  where $|\cM_n|$ denotes the number of messages included in $\cM_n$ and $\Pr\{\cdot\}$ denotes the probability of an event.  
  Also, the leaked information $\delta^E_n$ against Eve is measured as
    \begin{equation}
      \delta^E_n \equiv \frac{1}{|\cM_n|} \sum_{i \in \cM_n} D(P_n^{(i)}|| \pi_n), \label{delta}
    \end{equation}
  where $D(P_n^{(i)}||\pi_n)$ is the Kullback-Leibler distance \cite{CKbook} between the output probability distribution $P^{(i)}_n$ via the wiretapper channel due to message $i$ and 
  the target output probability distribution $\pi_n$ which is generated via the wiretapper channel due to an arbitrarily prescribed input distribution.  

  According to the theory of wiretap channel coding \cite{HES},
  there exists a code with length $n$ attaining the following upper bounds on the error probability $\vep^B_n$ and the leaked information $\delta^E_n$:
    \begin{align}
      \vep^B_n \leq 2 e^{-n \Fc}, \quad \delta^E_n \leq 2 e^{-n \Hc}, \label{upbound}
    \end{align}
  where the exponents $\Fc$ and $\Hc$ are referred to as the error exponent and the secrecy exponent defined as
    \begin{align}
      &\Fc \non \\
      & \quad \equiv \sup_{r \geq 0} \sup_{0 \leq \rho \leq 1} \left[\phiBr - \rho(R_B+R_E) \right], \label{relia1} \\
      &\Hc \non \\ 
      & \quad \equiv \sup_{r \geq 0} \sup_{0 < \rho < 1} \left[\phiEr + \rho R_E\right], \label{secre1}
    \end{align}
  respectively.  
  It is known that the error exponent $\Fc$ is a monotone strictly positive decreasing in $R_B + R_E < I(X;Y)$ and becomes $0$ for $R_B + R_E \ge I(X;Y)$.  
  Conversely, the secrecy exponent $\Hc$ is a monotone strictly positive increasing in $R_E > I(X;Z)$ and becomes $0$ for $ R_E \le I(X;Z)$.

  Here, $\phiBr$ in (\ref{relia1}) and $\phiEr$ in (\ref{secre1}) are functions of the given channels $W_B, W_E$ and the input probability $q$.  
  For the wiretap channel based on the OOK considered in this paper, these functions are given as in (\ref{phiXB}) and (\ref{phiXE}) at the top of the next page.
    \begin{figure*}
    \normalsize
    \setcounter{MYtempeqncnt}{\value{equation}}
    \setcounter{equation}{25}
    \begin{align}
      &\phiBr \non \\
      & \equiv -\log \left[ \left(qb_y^{\frac{1}{1+\rho}}e^{r\left(P - \frac{n^{\ast}_A h f_0 }{\Delta} \right)}
                            +(1-q)a_y^{\frac{1}{1+\rho}}e^{rP} \right)^{1+\rho} \right. 
                           + \left. \left(q(1-b_y)^{\frac{1}{1+\rho}}e^{r\left(P - \frac{n^{\ast}_A h f_0 }{\Delta} \right)}
                            +(1-q)(1-a_y)^{\frac{1}{1+\rho}}e^{rP} \right)^{1+\rho}\right] \label{phiXB}\\
      &\phiEr \non \\ 
      & \equiv -\log \left[ \left(qb_z^{\frac{1}{1-\rho}}e^{r\left(P - \frac{n^{\ast}_A h f_0 }{\Delta} \right)}
                            +(1-q)a_z^{\frac{1}{1-\rho}}e^{rP} \right)^{1-\rho} \right. 
                           + \left. \left(q(1-b_z)^{\frac{1}{1-\rho}}e^{r\left(P - \frac{n^{\ast}_A h f_0 }{\Delta} \right)}
                            +(1-q)(1-a_z)^{\frac{1}{1-\rho}}e^{rP} \right)^{1-\rho}\right] \label{phiXE}
    \end{align}
    \setcounter{MYtempeqncnt}{\value{MYtempeqncnt}}
    \hrulefill
    \vspace*{4pt}
    \end{figure*}
  The arbitrary constant $r \ge 0$ is optimized so that each exponent be maximized.  
  The authors of \cite{HES} have derived exponents for the wiretap channel with the auxiliary channel $P_{X|V}$.  
  In this paper, however, we only pay attention to the wiretap channel without $P_{X|V}$ for simplicity.

  \subsection{Code length dependence of error probability and leaked information}
  For a practical code of finite length $n$, the coding rate $R_B$ cannot be arbitrarily close to the secrecy capacity (or secrecy rate), as well as the error probability $\vep^B_n$ and the leaked information $\delta^E_n$ cannot be infinitesimally small.  
  In order to design the practical wiretap channel codes, the coding rate $R_B$ should be compromised to be much lower than the secrecy capacity, 
  and then the necessary code length $n$ for the required levels of $\vep^B_n$ and $\delta^E_n$ should be investigated.   
  This is actually the motivation to introduce the error exponent $\Fc$ and the secrecy exponent $\Hc$ \cite{gallager, csisres, hayashi}.   
  Although some previous studies (e.g. \cite{Wagner}) have revealed that the secrecy capacity can be asymptotically achieved with constructive codes, 
  the evaluation of both $\vep^B_n$ and $\delta^E_n$ for finite length codes has never been investigated to our best knowledge.   
  
  In Fig. \ref{zu52}, we show the upper bounds on $\vep^B_n$ and $\delta^E_n$ based on (\ref{upbound}) 
  choosing the case of the loss-independent region with $\alpha = 70$ dB and $\eta_{zy} = 0.9$.
  We again adopt a set of parameters as $P=10$ mW, $\lambda_y=10$ kcps, $\lambda_z=1$ cps, $\Delta=1$ ns, which are the same as in Section \ref{sec3}.  
  The secrecy rate $R_\mathrm{S}$ is $44.2$ Mbps and the optimum parameters are $n^{\ast}_A = 1.94 \times 10^7$, $q^{\ast} = 0.544$ (see Figs. \ref{zu35}, \ref{zu32}, and \ref{zu323}). 
  
  We fix the coding rate $R^{\ast}_B = 22.1$ Mbps as to be the half of the secrecy rate $R_{\mathrm{S}}$.  
  The solid line denotes the case of $R^\ast_E = 0.641$ Gbps which is set so that $F_c(q, R^\ast_B, R^\ast_E) = H_c(q, R^\ast_E) = 1.59 \times 10^{-4}$ as shown in table \ref{table1}.  
  As seen in this figure, both $\vep^B_n$ and $\delta^E_n$ begin to decrease rapidly over $n=10^4$ 
  and reach the standard error-free criterion $\vep^B_n < 10^{-9}$ and the leaked information criterion $\delta^E_n < 10^{-9}$ at around $n=10^5$, 
  which is the reasonable code length compared with the current technology.  
  In the standard channel coding without Eve, $\vep^B_n$ can be reduced arbitrarily by lowering the coding rate $R_B$ with fixing the code length.  
  However, in the wiretap channel coding, since $R_E$ should be kept larger than the mutual information $I(X;Z)$ for the secrecy against Eve, 
  it is not obvious whether there is a code of reasonable length $n$ which satisfies the required levels of both $\vep^B_n$ and $\delta^E_n$.  
  Fig. \ref{zu52} provides the significant knowledge on this point,
  namely, even for the relative transmittance $\eta_{zy} = 0.9$ which corresponds to the case where Eve can wiretap much power, 
  there is a practical code with sufficiently small $\vep^B_n$ and $\delta^E_n$.  

  In Fig. \ref{zu52}, the dashed line labeled with ``$R_E = 0.99R^\ast_E$" illustrates the case where $R_E$ is set to be $99\%$ of $R^\ast_E$.  
  As shown in table \ref{table1}, $\Fc$ increases compared to the case of $R^\ast_E$ because of its monotonicity in $R_B+R_E$.  
  This brings a decrease in $\vep^B_n$ as denoted by the arrow in the figure.
  On the other hand, $\Hc$ decreases because of its monotonicity in $R_E$ and $\delta^E_n$ increases.  
  As seen in the figure, $\vep^B_n$ reaches $10^{-9}$ around at $n=7 \times 10^4$, which is shorter than the case of $R^{\ast}_E$.  
  On the other hand, $\delta^E_n$ reaches only $10^{-1}$ with this code length.  
  In order to reach $\delta^E_n<10^{-9}$, a much longer code length of $n \ge 9 \times 10^{5}$ is required.  
  In contrast to the above case, the dotted line labeled with ``$R_E = 1.01R^\ast_E$" illustrates the case where $R_E$ is set to be $101\%$ of $R^\ast_E$.  
  In this case, $\vep^B_n$ increases whereas $\delta^E_n$ decreases as shown in the figure.  

  Intuitive examples of the above discussion are as follows;
  in order to relax the implementation cost of codes, one may wish to change the criteria for the secrecy according to the level of confidentiality of information.  
  In the opposite case, more secure codes may be required to establish secure links leaving the complexity of implementation out of consideration.  
  The discussion in this subsection provides the quantification of such an adaptive change of the performances.  
  In other words, we characterize another clue for controlling the tradeoff between performance and code length via the upper bounds in (\ref{upbound}), 
  which is more practical than other examples of tradeoff relation provided in \cite{HES,chou}.  
  
\section{Conclusion} \label{conclu}
  In this paper, we have studied the performance of physical layer security of FSO communications based on the OOK modulation with linear attenuation and background noises,  
  using the secrecy capacity and the code length dependence of the error probability and the leaked information as performance metrics. 
  Although we have mainly focused on the idealistic setting, i.e., without fading, 
  we have numerically shown that the global scale network with ITS would be potentially realized by wiretap channel coding  with currently available technologies and 
  there exists a wiretap channel code of a practical length.

  We have numerically investigated the secrecy rates and the secrecy capacity and clarified its unique features as follows; 
  (a) unless the transmission power is regulated optimally, these quantities dramatically drop in the small attenuation region and
  (b) transmission distance of our proposed method can be much longer than that of QKD even when Eve can obtain $99\%$ as much the fraction of power as Bob.  
  We have also shown that the transmission distance can be extended by introducing the auxiliary random variable $V$ at Alice \cite{csiskor}
  if the wiretap channel is not more capable.  
  The random additional pulses resulting from the auxiliary channel $P_{X|V}$ play an essential role in deceiving Eve when the SNR at Bob is worse,
  which implies the similarity to the decoy method employed in QKD.  
  This physical implication of the effect of the auxiliary random variable has not been explicitly demonstrated so far.  

  Further, on the basis of the past theoretical study \cite{HES}, we have introduced the error exponent $\Fc$ and the secrecy exponent $\Hc$ for our proposed method.  
  We have provided the characterization of such exponents in terms of the code length dependence of the error probability $\vep^B_n$ and the leaked information $\delta^E_n$.  
  The code length dependence of $\vep^B_n$ and $\delta^E_n$ provides 
  (a) the evaluation of $\vep^B_n$ and $\delta^E_n$ for practical codes of finite length and, 
  (b) the necessary code length to satisfy the required levels of both $\vep^B_n$ and $\delta^E_n$.  
  Our calculation has indicated the existence of a practical code with the reasonable length and the sufficient performance 
  even for the case where Eve can obtain $90\%$ as much power as Bob.  
  
  There might be many interesting problems left open. We mention two of them. First, our analysis
  should be extended to include the fading effect. The received signal intensity through a typical 
  FSO channel fluctuates in a time scale of millisecond due to atmospheric scintillation.  A
  straightforward way is to model this fluctuation by renormalizing the noise variance in a log-normal
  fading distribution, which leads to the degradation of overall performances. A more sophisticated
  approach is an adaptive scheme. If the CSI can be estimated by Alice, the transmission power
  can be allocated opportunistically to the instantaneous fading realizations for which Eve obtains
  a lower instantaneous SNR than that of Bob. As a result, strictly positive secrecy rates are
  achievable even if, on average, Eve obtains a better SNR than that of Bob \cite{BBbook, Bloch}. 
  However, this adaptive scheme requires a fast feedforward mechanism in the millisecond time scale, and remains a challenge.  
  
  Second and the last, multiple colluding eavesdroppers are a likely risk in an FSO link.  
  One can easily imagine that multiple drones tap various places in the FSO link, and collude for getting information.  
  Countermeasures should not be simple, and be sought from the viewpoint not only of coding schemes but also of system level solution like monitoring and alarming functions. 
  
  In spite of such a challenging problem to which we should address in the future, 
  we believe that the potential performances of physical layer security of FSO communications presented in this paper 
  provide insight into a new direction for secure communications.  
  For example, it is noteworthy that performances of physical layer security of FSO channels and QKD are regarded as complementary technologies in the sense of the tradeoff between security level and usability.  
  Thus, they will eventually be integrated to realize high capacity optical communications with ITS and such a combination should provide the new paradigm of secure communications.

\section*{Acknowledgement}
  This work was funded by ImPACT Program of Council for Science, Technology and Innovation (Cabinet Office, Government of Japan).

\bibliography{IEEEabrv,Reference_PJ-003559-2015}

\end{document}